\documentclass[useAMS,usenatbib]{mn2e_mod}
\usepackage{epsfig,amsmath,amssymb,amsfonts,natbib,color,url}  

\usepackage{amsmath,amssymb,amsfonts,natbib}
\usepackage {graphicx,color}
\usepackage {txfonts }
\usepackage {natbib  }
\usepackage {float   }
\usepackage {enumerate}
\usepackage {subfigure  }

\newcommand{\revised}[1]{{#1}} 
\newcommand{\rev}[1]{{#1}}

\newcommand{\e}[1]{\ensuremath{\times 10^{#1}}}
\newcommand{\sub}[1]{_{\rm{#1}}}
\newcommand{\xdust}{\ensuremath{x\sub{dust}} }

\newcommand{\xmax}{\ensuremath{x\sub{dust,\,max}}}
\newcommand{\xamp}{\ensuremath{x\sub{dust,\,amp}}}

\def\gm{{\rm\,g}}

\title{Catastrophic Evaporation of Rocky Planets}

\author[D.~Perez-Becker \&\,E.~Chiang]{Daniel~Perez-Becker$^{1}$\footnotemark[1] and\,Eugene~Chiang$^{2}$\\
 \\
  $^1$Department of Physics, University of California at Berkeley, 366 LeConte Hall, Berkeley CA 94720-7300, USA\\
  $^2$Departments of Astronomy and of Earth and Planetary Science, University of California at Berkeley, Hearst Field Annex B-20, Berkeley CA 94720-3411, USA\\}

\date{Submitted: \today}

\begin{document}
\maketitle
\begin{abstract}
  Short-period exoplanets can have dayside surface temperatures
  surpassing 2000 K, hot enough to vaporize rock and drive a thermal
  wind. Small enough planets evaporate completely.  We construct a
  radiative-hydrodynamic model of atmospheric escape from strongly
  irradiated, low-mass rocky planets, accounting for dust-gas energy
  exchange in the wind. Rocky planets with masses $\lesssim 0.1
  M_{\earth}$ (less than twice the mass of Mercury) and surface
  temperatures $\gtrsim 2000$ K are found to disintegrate entirely in
  $\lesssim$ 10 Gyr.  When our model is applied to Kepler planet
  candidate KIC 12557548b --- which is believed to be a rocky body
  evaporating at a rate of $\dot{M} \gtrsim 0.1 \,M\sub{\earth}$/Gyr
  --- our model yields a present-day planet mass of $\lesssim 0.02
  \,M\sub{\earth}$ or less than about twice the mass of the Moon. Mass
  loss rates depend so strongly on planet mass that bodies can reside
  on close-in orbits for Gyrs with initial masses \revised{comparable
    to or less than} that of Mercury, before entering a final
  short-lived phase of catastrophic mass loss (which KIC 12557548b has
  entered). Because this catastrophic stage lasts only \revised{up to}
  a few percent of the planet's life, we estimate that for every
  object like KIC 12557548b, \revised{there should be 10--100 close-in
    quiescent progenitors with sub-day periods whose hard-surface
    transits may be detectable by \textit{Kepler} --- if the progenitors
    are as large as their maximal, Mercury-like sizes (alternatively,
    the progenitors could be smaller and more numerous)}. According to
  our calculations, KIC 12557548b may have lost $\sim$70\% of its
  formation mass; today we may be observing its naked iron core.

\end{abstract}

\label{firstpage}
\footnotetext[1]{e-mail: \rm{\url{perez-becker@berkeley.edu}.}}

\section{INTRODUCTION} 
\label{sec_intro}

Atmospheric escape shapes the faces of planets. Mechanisms for mass
loss vary across the Solar System. \revised{Planets lacking magnetic
  fields can be stripped of their atmospheres by the Solar wind: the
  atmospheres of both Mercury and Mars are continuously eroded and
  refreshed by Solar wind ions
  \citep{Potter:1990p17035,Potter:1997p16997,Lammer:1997p16995,Bida:2000p17040,Killen:2004p17021,Jakosky:2001p17058}.}
Venus demonstrates the extreme sensitivity of atmospheres to stellar
insolation.  Although its distance to the Sun is only 30\% less than
that of the Earth, Venus has lost its water to hydrodynamic escape
powered by Solar radiation
\citep{Kasting:1983p16892,Zahnle:1986p16926}. Impacts with comets and
asteroids can also purge planets of their atmospheres, as has thought
to have happened \revised{to some extent} on Mars
\citep{Melosh:1989p17048, Jakosky:2001p17058} and some giant planet
satellites \citep{Zahnle:1992p17063}. An introductory overview of
atmospheric escape in the Solar System is given by
\citet{Catling:2009p16873}.

Smaller bodies lose their atmospheres more readily because of their lower surface gravities. Bodies closer to their host star also lose more mass because they are heated more strongly. Extrasolar planets on close-in orbits are expected to have highly processed atmospheres. In extreme cases, stellar irradiation might even evaporate planets in their entirety --- comets certainly fall in this category.

Hot Jupiters are the best studied case for atmospheric erosion in extrasolar planets (for theoretical models, see \citealt{Yelle:2004p16652, Yelle:2006p16662, Tian:2005p16700, GarciaMunoz:2007p3342, Holmstrom:2008p16863, MurrayClay:2009p16619, Ekenback:2010p16864, Tremblin:2013p16188}). For these gas giants, mass loss is driven by stellar ultraviolet (UV) radiation which photo-ionizes hydrogen and imparts enough energy per proton to overcome the planet's large escape velocity ($\sim$60 km/s). 
Mass loss for hot Jupiters typically occurs at a rate of $\dot{M} \sim 10^{10}$--$10^{11}$ g/s so that $\sim$1\% of the planet mass is lost over its lifetime \citep{Yelle:2006p16662, GarciaMunoz:2007p3342, MurrayClay:2009p16619}. Although this erosion hardly affects a hot Jupiter's internal structure, the outflow is observable.  Winds escaping from hot Jupiters have been spectroscopically detected by the \textit{Hubble Space Telescope (HST)}. During hot Jupiter transits, several absorption lines (H$\,$\textsc{i}, O$\,$\textsc{i}, C$\,$\textsc{ii}, Mg$\,$\textsc{ii}, and Si$\,$\textsc{iii}) deepen by $\sim$2--10\% \citep{VidalMadjar:2003p16627,VidalMadjar:2004p16629, VidalMadjar:2008p16632, BenJaffel:2007p16634, BenJaffel:2008p16635, Fossati:2010p16641, LecavelierDesEtangs:2010p16644, Linsky:2010p16647, LecavelierDesEtangs:2012p16826}.

Because of their lower escape velocities, lower mass super-Earths should have their atmospheres more significantly sculpted by evaporation. \citet{Lopez:2012p16574} found that hydrogen-dominated atmospheres of close-in super-Earths could be stripped entirely by UV photoevaporation (e.g., Kepler-11b). %

At even lower masses and stellocentric distances, planets might
evaporate altogether. \revised{When dayside temperatures surpass
  $\sim$2000 K, rocks, particularly silicates and iron (and to a
  lesser extent, Ti, Al, and Ca), begin to vaporize.} If the planet
mass is low enough, thermal velocities of the high-metallicity vapor
could exceed escape velocities. Under these circumstances, mass loss
is not limited by stellar UV photons, but is powered by the incident
bolometric flux.

\subsection{KIC 12557548b: A Catastrophically Evaporating Planet} \label{sec_intro_KIC}

In fact, the \textit{Kepler} spacecraft may have enabled the discovery of an evaporating, close-in rocky planet near the end of its life. As reported by \citet{Rappaport:2012p14995}, the K-star KIC 12557548 dims every 15.7 hours \revised{by a fractional amount $f$} that varies erratically from a maximum of $f \approx 1.3\%$ to a minimum of $f \lesssim 0.2\%$. These occultations are thought to arise from dust embedded in the time-variable outflow emitted by an evaporating rocky planet --- hereafter KIC 1255b. \revised{Because {\it Kepler} observes in the broadband optical, the obscuring material cannot be gas which absorbs only in narrow lines; it must take the form of dust which absorbs and scatters
efficiently in the continuum. As we will estimate shortly, the amount of dust lost by the planet
is prodigious.}

From the $\sim$2000 K surface of the planet is launched a thermal
(``Parker-type'') wind out of which dust condenses as the gas expands
and cools. \rev{The composition of the high-metallicity wind reflects
  the composition of the evaporating rocky planet (see, e.g.,
  \citealt{Schaefer:2009p15499}): the wind may consist of Mg, SiO, O,
  and O$_2$ --- and Fe if it has evaporated down to its iron core.}
Stellar gravity, together with Coriolis and radiation-pressure forces,
then shape the dusty outflow into a comet-like tail. The peculiar
shape of the transit light curve of KIC 1255b supports the
interpretation that the occultations are caused by a dusty
tail. First, the light curve evinces a flux excess just before
ingress, which could be caused by the forward scattering of starlight
toward the observer. Second, the flux changes more gradually during
egress than during ingress, as expected for a trailing tail. Both
features of the light curve and their explanations were elucidated by
\citet{Rappaport:2012p14995}, and further quantified by
\citet{Brogi:2012p15046} who presented a phenomenological
light-scattering model of a dusty tail.\footnote{ \rev{ Alternative
    ideas might include Roche lobe overflow from a companion to the
    star (c.f.~\citealt{Lai:2010p17222}). A problem with this scenario
    is that the mass transfer accretion stream, funneled through the
    L1 Lagrange point, would lead the companion in its orbit, not
    trail it as is required by observation. By comparison, in our
    dusty wind model, there is no focussing of material at L1 because
    the wind is not in hydrostatic} \rev{equilibrium} \rev{and
    therefore does not conform to Roche lobe equipotentials. Moreover,
    stellar radiation pressure, usually neglected in standard
    Roche-overflow models, is crucial for diverting the flow so that
    it trails the planet in its orbit (see also \citealt{Mura:2011p17293}).}
  \rev{Another alternative involves bow shocks of planets passing
    through stellar coronae (\citealt{Vidotto:2010p17160,
      Vidotto:2011p17172, Vidotto:2011p17167, Llama:2011p17173,
      Tremblin:2013p16188})} \rev{which} \rev{may explain the
    ingress-egress asymmetry observed for the exoplanet WASP-12b
    (\citealt{Fossati:2010p16641}). Bow shocks between the planetary
    wind and the stellar wind are also expected to occur for KIC
    1255b, but would be hard to detect through {\it Kepler}'s
    broadband filter since the shocked gas alone would absorb only in
    narrow absorption lines.  At the low gas densities characterizing
    exoplanet winds, only dust can generate the large continuum
    opacities required by the deep transits observed by {\it
      Kepler}. \citet{Rappaport:2012p14995} reviews these and
    additional arguments supporting the dust tail model.  A similar
    interpretation was proposed for the light curve of beta Pictoris
    (\citealt{LecavelierDesEtangs:1997p17174};
    \citealt{Lamers:1997p17184}).}}

Spectra of KIC 12557548 taken using the Keck Telescope reveal no
radial velocity variations down to a limiting accuracy of $\sim$100
m/s (A. Howard and G. Marcy 2012, personal communication). A deep Keck
image also does not reveal any background blended stars. These null
results are consistent with the interpretation that KIC 1255b is a
small, catastrophically evaporating planet.

\revised{The characteristic mass loss rate for KIC 1255b --- a quantity that we will
reference throughout this paper --- can be estimated
from the observations as follows. First 
compute the total mass in grains required
to absorb/scatter a fraction $f$ of the starlight, assuming that 
the dust is optically thin (or marginally so). Out of the total area $\pi R_{\ast}^2$
presented by the stellar face, take
the dust cloud to cover an area $A$ 
and to have a characteristic line-of-sight optical depth $\Sigma_{\rm d} \kappa_{\rm d}$,
where $\Sigma_{\rm d}$ is the grain mass per unit area (on the sky)
and $\kappa_{\rm d}$ is the grain opacity. Then
\begin{equation}
f = \frac{A}{\pi R_{\ast}^2} \Sigma_{\rm d} \kappa_{\rm d} \,.
\end{equation}
Assume further that 
grains can be modeled as a monodispersion of spheres with radii $s$
and internal \rev{material} density $\rho_{\rm int}$ so that
\begin{equation}
\kappa_{\rm d} = \frac{3}{4 \rho_{\rm int} s} \,.
\end{equation}
The total mass in dust covering the star is then
\begin{equation}
M_{\rm d} = \Sigma_{\rm d} A = \frac{4\pi}{3} f \rho_{\rm int} R_{\ast}^2 s  \,.
\end{equation}
To obtain a mass loss rate, divide this mass by the orbital period
of 15.7 hours on the grounds that transit depths change by up to an order
of magnitude from orbit to orbit. For $f = 1\%$ --- when KIC 1255b is in a state of relatively high mass loss --- the resultant mass loss rate in grains is
\begin{equation}
\dot{M}\sub{1255b,\, dust} \sim 0.4\, \left( \frac{R_{\ast}}{0.65 R_\odot} \right)^2 \left( \frac{f}{0.01} \right) \left( \frac{s}{0.5 \,\mu{\rm m}} \right) \left( \frac{\rho_{\rm int}}{1 \gm/{\rm cm}^{3}} \right)  M\sub{\earth}\, {\rm Gyr}^{-1} \,.
\label{eq_1255b_dust}
\end{equation}
This is the mass loss rate in dust only, \rev{by construction}.
The gas-to-dust ratio by mass, $\eta$, 
must be at least on the order of unity if outflowing gas is to carry, by Epstein drag,
embedded grains out of the gravity well of the planet \citep{Rappaport:2012p14995}.
Henceforth we will adopt as our fiducial total mass loss rate (for the relatively high value of $f = 1\%$):
\begin{equation}
\dot{M}\sub{1255b} = (1 + \eta) \dot{M}\sub{1255b,\,dust} \sim 1\, M\sub{\earth}\,{\rm Gyr}^{-1} \,.
\label{eq_1255b}
\end{equation}
The total mass loss rate $\propto \rho_{\rm int} s (1 + \eta)$.  Given
the uncertainties in these three factors \rev{(some rough guesses: $\rho_{\rm int} \gtrsim
  0.5$ g cm$^{-3}$ for possibly porous grains; $s \gtrsim 0.1 \, \mu{\rm m}$;
  $\eta \gtrsim 1$)}, the total mass loss rate could take any value
$\dot{M}\sub{1255b} > 0.1 \,M\sub{\earth}$ / Gyr (but see also our
Figures \ref{fig_F} and \ref{fig_H} which place upper bounds on mass
loss rates).  }

\citet{Rappaport:2012p14995} claimed that the present-day mass of KIC
1255b is $\sim$0.1 $M\sub{\earth}$. They estimated, to
order-of-magnitude, the maximum planet mass that could reproduce the
observed mass loss rate given by equation (\ref{eq_1255b}). Here we
revisit the issue of planet mass --- and the evaporative lifetime it
implies --- with a first-principles solution of the hydrodynamic
equations for a planetary wind, including an explicit treatment of
dust-gas thermodynamics. We will obtain not only improved estimates
of the present-day mass, but also full dynamical histories of KIC
1255b.

\subsection{Plan of This Paper}

We develop a radiative-hydrodynamic wind model to study mass loss of close-in rocky planets with dayside temperatures high enough to vaporize rock. %
Although our central application will be to understand KIC 1255b --- both today and in the past --- the model is general and can be easily be modified for parameters of other rocky planets. %

The paper is structured as follows. In \S\ref{sec_model} we present the 
 model that computes $\dot{M}$ as a function of planet mass $M$.  We give results in \S\ref{sec_results}, including estimates for the \revised{maximum} present-day mass and the \revised{maximum} formation mass of KIC 1255b. A wide-ranging discussion of the implications of our results --- including, e.g., an explanation of why mass loss in this context does not obey the often-used energy-limited mass loss formula (cf.~\citealt{Lopez:2012p16574}), as well as the prospects of detecting close-in Mercuries and stripped iron cores --- is given in \S\ref{sec_discussion}. Our findings are summarized in \S\ref{sec_conclusions}.

\section{THE MODEL}
\label{sec_model}
We construct a 1D model for the thermally-driven atmospheric mass loss from a close-in rocky planet, assuming that it occurs in the form of a transonic Parker wind. The atmosphere of the planet consists of a \revised{high-metallicity} gas created by the sublimation of silicates (and possibly iron, as we discuss in \S\ref{sec_iron}) from the planetary surface. %
We focus on the flow that streams radially toward the star from the substellar point on the planet. The substellar ray is of special interest because the mass flux it carries will be maximal insofar as the substellar point has the highest temperature --- and thus the highest rate of rock sublimation --- and insofar as stellar gravity will act most strongly to accelerate gas away from the planet. 

As the \revised{high-metallicity} gas flows away from the surface of the planet, it will expand and cool, triggering the condensation of dust grains. Dust grains can affect the outflow in several ways: (a) latent heat released during the condensation of grains will heat the gas; (b) continuously heated by starlight, grains will transfer their energy to the gas via gas-grain collisions; and (c) absorption/scattering of starlight by grains will reduce the stellar flux reaching the planet, reducing the surface temperature and sublimation rate. \revised{In this work, we do not account for the microphysics of grain condensation, as cloud formation is difficult to calculate from first principles (see, e.g.,
the series of five papers beginning with \citealt{Helling:2001p17233};
see also \citealt{Helling:2008p17236} and \citealt{Helling:2008p17230}).}
Instead, we treat the dust-to-gas mass fraction $\xdust$ as a free
function and explore how our results depend on this function.  \revised{In our
1-fluid model, dust grains are perfectly entrained in the gas flow, an
assumption we test \textit{a posteriori} (\S\ref{sec_1fluid}).}

We begin in \S\ref{sec_atmosphere} by evaluating thermodynamic variables at the surface of the planet --- the inner boundary of our model. In \S\ref{sec_hydro} we introduce the 1D equations of mass, momentum, and energy conservation. In \S\ref{sec_relax} we explain the relaxation method employed to solve the conservation laws.

\subsection{Base Conditions of the Atmosphere} \label{sec_atmosphere}

\begin{figure}%
\includegraphics[width=\linewidth]{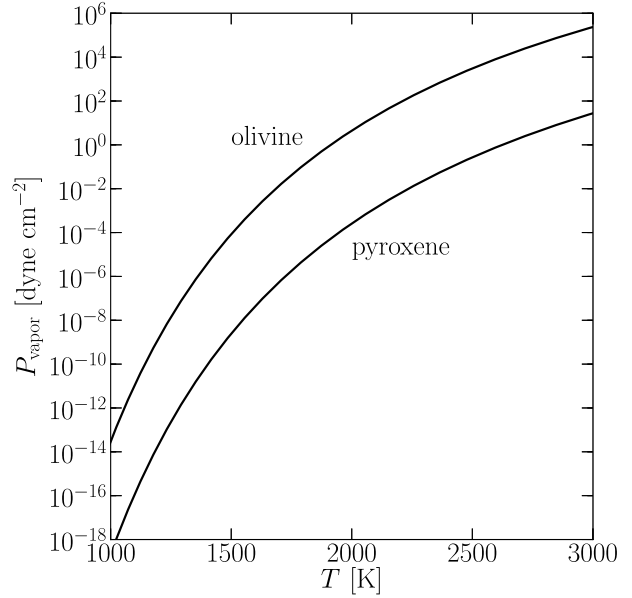}
\caption{
Equilibrium vapor pressures of pyroxene and olivine vs.~temperature, obtained from equation~(\ref{eq_pvap}). }
\label{fig_A}
\end{figure}

Dayside planet temperatures peak at approximately
\begin{equation}
	T\sub{surface}\approx T\sub{*}\sqrt{\frac{R\sub{*}}{a}} e^{-\tau\sub{surface}/4} \approx 2150\,\, e^{-\tau\sub{surface}/4}\,\,\rm{K},
	\label{eq_Tsurf}
\end{equation}
where for our numerical evaluation we have used parameters appropriate to KIC 12557548 (orbital semimajor axis $a \approx 0.013$ AU; effective stellar temperature $T\sub{*} \approx 4450$ K; and stellar radius $R\sub{*} \approx 0.65\, R_{\sun}$;  \citealt{Rappaport:2012p14995}). Here $\tau\sub{surface}$ is the optical depth between the planet's surface and the star due to absorption/scattering by dust grains.

We expect rocky extrasolar planets to consist of the same silicates predominantly found in the mantle of the Earth: pyroxene ([Fe,Mg]SiO$_3$) and olivine ([Fe,Mg]$_2$SiO$_4$). At the estimated $T\sub{surface}$, these silicates will vaporize and form an atmosphere whose equilibrium vapor pressure can be described by 
\begin{equation} 
P\sub{vapor}=\exp \left(-\frac{m L\sub{sub}}{kT\sub{surface}} + b \right),
\label{eq_pvap}
\end{equation}
where $m$ is the mass of either a pyroxene or olivine molecule, $k$ is
Boltzmann's constant, $L\sub{sub}$ is the latent heat of sublimation,
and $b$ is a constant. We follow \citet{Kimura:2002p14382} in our
choice of parameters for equation (\ref{eq_pvap}). For olivine,
$m=169$ m$\sub{H}$ (Mg$\sub{1.1}$Fe$\sub{0.9}$SiO$\sub{4}$) where
m$\sub{H}$ is the mass of atomic hydrogen, $L\sub{sub}=3.21\e{10}$ 
erg~g$^{-1}$, and $e^{b}= 6.72\e{14}$ dyne cm$^{-2}$; these values were
found experimentally by evaporating forsterite (Mg$_2$SiO$_4$), an
end-member of olivine \citep{Nagahara:1994p15612}. For pyroxene,
$m=60$ m$\sub{H}$, $L\sub{sub}=9.61\e{10}$ erg g$^{-1}$, and $e^{b}=
3.13\e{11}$ dyne cm$^{-2}$; these values were determined by
\citet{Hashimoto:1990p15508} for SiO$_2$. \citet{Kimura:2002p14382}
argue that the parameters for SiO$_2$ are appropriate for pyroxene
because evaporation experiments of enstatite (MgSiO$_3$) show that
SiO$_2$ escapes preferentially, leaving the mineral with a
polycrystalline forsterite coating \citep{Tachibana:2002p15629}. We
plot equation (\ref{eq_pvap}) for both olivine and pyroxene in Figure
\ref{fig_A}.

According to \citet{Schaefer:2009p15499}, 
the atmosphere of a hot rocky planet at temperatures similar to $T\sub{surface}$ is primarily composed of Mg, SiO, O, and O$_2$. In our model we choose a mean molecular weight for the atmospheric gas of 
\begin{equation}
\mu=30 \,\,\rm{m}\sub{H}, 
\end{equation}
similar to the average molecular mass of Mg and SiO.
Additionally, we set the adiabatic index of the gas to be $\gamma \equiv c\sub{P}/c\sub{V}=1.3$, appropriate for diatomic gases at high temperatures (e.g., $\gamma\sub{H_2}^{T=2000\rm{K}}=1.318$, \citealt{Lange:1967p0001}). The gas density at the surface of the planet is  
\begin{equation}
\rho\sub{vapor}=\frac{\mu P\sub{vapor}}{kT\sub{surface}}.
\end{equation}

\subsection{Equations Solved} \label{sec_hydro}
In our 1D hydrodynamic model, we solve the equations for mass, momentum, and energy conservation of the gas, assuming a steady flow. From mass \revised{conservation},   
\begin{equation}
	\frac{\partial}{\partial r}(r^2 \rho v)=0,
\label{eq_mass}
\end{equation}
where $r$ is the radial distance from the center of the planet, and $\rho$ and $v$ are the density and velocity of the gas. In the frame rotating at the orbital angular frequency of the planet, momentum conservation --- with the Coriolis force omitted --- implies 
\begin{equation}
	\rho v \frac{\partial v}{\partial r} = -\frac{\partial P}{\partial r} -\frac{GM\rho}{r^2} + \frac{3 G M\sub{\star}\rho r}{a^3},
\label{eq_momentum}
\end{equation}
where $G$ is the gravitational constant, $P=\rho kT/\mu$ is the gas
pressure, $T$ is the gas temperature, and $M$ and $M\sub{*}=0.7
M_{\sun}$ are the masses of the planet and the host star (KIC
12557548), respectively. The last term on the right-hand side of
eq. (\ref{eq_momentum}), which we refer to as the tidal gravity term,
is the sum of the centrifugal force and the gravitational attraction
from the star. In deriving eq. (\ref{eq_momentum}), we have neglected
terms of order $(r/a)^2$. We do not include the contribution of the
Coriolis force because its magnitude will only be comparable to the
gravitational attraction of the planet when the gas moves at bulk
speeds comparable to the escape velocity of the planet at its Hill
sphere (a.k.a. the Roche lobe). This generally occurs near the outer
boundary of our calculation, which is set by the location of the sonic
point. Although the sonic point radius and Hill sphere radius are
distinct quantities, we will find in practice for our models that they
are close to one another, which is not surprising since the wind most
easily accelerates to supersonic velocities where the effective
gravity is weakest --- i.e., near the Hill sphere.

The steady-state energy conservation equation, $\nabla \cdot (\rho u \textbf{v})= -P\nabla \cdot \textbf{v} + \Gamma$, together with equation (\ref{eq_mass}) and the specific internal energy $u=kT/[(\gamma-1)\mu]$, yields

\begin{equation}
	\rho v \frac{\partial }{\partial r} \left[ \frac{kT}{(\gamma-1)\mu}\right]= \frac{kT v }{\mu}\frac{\partial \rho}{\partial r} + \Gamma\sub{col} +\Gamma\sub{lat}. 
\label{eq_energy}
\end{equation}
The left hand side of equation (\ref{eq_energy}) tracks changes in internal energy, while the terms on the right hand side track cooling due to $PdV$ work, heating from gas-grain collisions, and latent heat released from grain condensation. We have omitted conduction because we found it to be negligible compared to all other terms.  

\revised{
For gas heating by dust-gas collisions, we assume that each
collision between a gas molecule and a dust grain transfers
an energy $k(T\sub{dust} - T)$ to the gas molecule.
Then the rate of gas heating per unit volume is}
\begin{equation}
\Gamma\sub{col}=  \frac{3 \xdust }{4 s \rho\sub{int}}  \frac{k}{\mu} \left(T\sub{dust}-T \right) \rho^2 c\sub{s},
\end{equation}
where $x\sub{dust}(r) \equiv \rho_{\rm dust}/\rho$ is the spatially varying dust-to-gas mass ratio, $s=1$ $\mu$m is the assumed grain size,  $\rho\sub{int}= 3$ g cm$^{-3}$ is the grain bulk density, and 
\begin{equation}
c\sub{s}= \sqrt{\frac{\gamma k T}{\mu}}
\end{equation}
is the sound speed of the gas. \revised{As discussed in
  \S\ref{sec_1fluid}, our assumption that micron-sized grains are
  entrained in the flow is valid for the lowest mass planets we consider
  (a group that includes, as best we will gauge, KIC 1255b)
  but not the highest mass planets.}

We adopt
\begin{equation}
T\sub{dust}= T\sub{*}\sqrt{\frac{R\sub{*}}{a}} e^{-\tau/4}
\end{equation}
for simplicity, so that $T\sub{dust}$ equals $T\sub{surface}$ at the surface of the planet. The dust optical depth to starlight from $r$ to the star is
\begin{equation}
\tau = \frac{3} {4 s \rho\sub{int}} \int_{r}^{r\sub{s}} \rho x\sub{dust} dr + \tau\sub{s}.
\label{eq_tau}
\end{equation}
The constant $\tau\sub{s}$ accounts for the optical depth from the outer boundary of our calculation --- the sonic point radius $r\sub{s}$ --- to the star. We arbitrarily set 
\begin{equation}
\tau\sub{s}=0.01.
\end{equation} 
Our results are insensitive to our assumed value of $\tau\sub{s}$ as long as it is $\ll 1$. The optical depth at the planet's surface is $\tau\sub{surface} \equiv \tau(r=R)$, where $R$ is the planetary radius, calculated for a given
$M$ by assuming a constant bulk density of 5.4 g/cm$^{3}$, the mean density of Mercury.

Latent heat from grain formation is released at a rate given by 
\begin{equation}
\Gamma\sub{lat}= L\sub{sub} \rho v \frac{dx\sub{dust}}{dr},
\end{equation}
where we have used $L\sub{sub}$ appropriate for pyroxene, as it
\rev{is more thermally stable} than olivine (see Figure \ref{fig_A}; in any case, $L\sub{sub}$
for olivine is not that different).
 
Because we do not model dust condensation from first principles, the
dust-to-gas mass ratio $\xdust (r)$ is a free function, which we
parametrize as follows:
\begin{eqnarray}
\log[\xdust(r)] &=& \log(x\sub{dust,\,max}) - \log(x\sub{dust,\,amp}) \,\cdot \cr & & \left\{ 1 - \arctan\left[\frac{8(r-R)}{3(r\sub{s}-R)}\right] \left[ \arctan \left( \frac{8}{3}\right)\right]^{-1} \right\}. 
\label{eq_xdust}
\end{eqnarray}
The function $\xdust$ increases monotonically with $r$ to a maximum
value of $x\sub{dust,\,max}$, starting from a minimum value of
$x\sub{dust,\,max} / x\sub{dust,\,amp}$.  We have two reasons for
choosing $\xdust$ to increase rather than decrease with increasing
height. The first is physical: as the wind launches into space, it
expands and cools, allowing the metal-rich vapor to more easily
saturate and condense (see also \S\ref{sec_condensation}). The second
is motivated by observations: on length scales on the order of the
stellar radius, far from the planet, the wind must have a relatively
high dust content so as to produce the deep transits observed by {\it
  Kepler}. \rev{Of course, neither of these reasons constitutes a
  first-principles proof that $\xdust$ actually does increase with $r$. Indeed the
  increasingly low density of the wind may frustrate condensation. A
  physics-based calculation of how grains may (or may not) form in
  the wind is sorely needed (\S\ref{sec_condensation}); our present
  study merely parameterizes this difficult problem.  }

As dust condenses from gas and $\xdust$ rises, the gas density $\rho$ must
fall by mass conservation.  For simplicity we neglect the dependence
of $\rho$ on $\xdust$ and so restrict our analysis to
$x\sub{dust,\,max} \ll 1$; in practice, the maximum value of
$x\sub{dust,\,max}$ that we consider is $10^{-0.5} \approx 0.32$.

\subsection{Relaxation Code} \label{sec_relax}

The structure of the wind is found by the simultaneous solution of equations (\ref{eq_mass}), (\ref{eq_momentum}), (\ref{eq_energy}), \& (\ref{eq_tau}) with the appropriate boundary conditions. This system constitutes a two-point boundary value problem, as the boundary conditions for $\rho$ and $T$ are defined at the base of the flow, while those for $v$ and $\tau$ are defined at the sonic point. 

Although generally more complicated, relaxation codes are preferred over shooting methods when solving for the transonic Parker wind. This is because for each transonic solution there are an infinite number of ``breeze solutions'' where the bulk speed of the wind never reaches supersonic velocities. It is easier to begin with an approximate solution that already crosses the sonic point and to then refine this solution, than to exhaustively shoot in multidimensional space for the sonic point. \citet{MurrayClay:2009p16619} used a relaxation code to find the transonic wind from a hot Jupiter. Here we follow their methodology; we develop a relaxation code based on Section 17.3 of \citet{Press:1992p15493}.

\subsubsection{Finite Difference Equations}

We set up a grid of $N=100$ logarithmically spaced radii that run from the base of the flow located at the surface of the planet ($r=R$) to the sonic point ($r=r\sub{s}$). We replace our system of differential equations by a set of finite-difference relations. For $\rho$ and $T$:

\begin{eqnarray}
E_{1,j} &\equiv & \Delta_{j}\rho - \frac{d\rho}{dr} \Delta_{j} r\cr
 &=&  \Delta_{j}\rho + \rho \left(\frac{2}{r} + \frac{1}{v}\frac{dv}{dr}   \right)\Delta_{j} r=0,
\label{eq_E1}
\end{eqnarray}

\begin{eqnarray}
E_{2,j} &\equiv & \Delta_{j} T - \frac{dT}{dr} \Delta_{j} r\cr
 &=&  \Delta_{j}T -  (\gamma-1) \left[ \frac{T}{\rho} \frac{d\rho}{dr} + \frac{\mu}{k \rho v} (\Gamma\sub{col}+\Gamma\sub{lat}) \right]  \Delta_{j} r=0,
\label{eq_E2}
\end{eqnarray}
where finite differences are calculated upwind ($\Delta_j x \equiv x_j - x_{j-1}$), as $\rho$ and $T$ (and by extension $E_{1}$ and $E_{2}$) have boundary conditions defined at the surface of the planet ($j=0$). Following \citet{Press:1992p15493}, we average across adjacent grid points when evaluating variables in (\ref{eq_E1}) and (\ref{eq_E2}), e.g., $\rho \equiv (\rho_{j-1}+\rho_j)/2$. The finite-difference relations for $v$ and $\tau$ are:

\begin{eqnarray}
E_{3,j} &\equiv & \Delta_{j} v - \frac{dv}{dr} \Delta_{j} r\cr
 &=&  \Delta_{j}v -  \frac{v}{v^2-\gamma kT/\mu} \Biggl[  \frac{2\gamma kT}{\mu r} - \frac{\gamma-1}{\rho v}(\Gamma\sub{col}+\Gamma\sub{lat}) \cr
&  & -GM/r^2+ 3GM_{*}r/a^3 \Biggr] \Delta_{j} r=0,
\label{eq_E3}
\end{eqnarray}

\begin{eqnarray}
E_{4,j} &\equiv & \Delta_{j} \tau - \frac{d\tau}{dr} \Delta_{j} r\cr
 &=&  \Delta_{j}\tau + \frac{3}{4 s \rho\sub{int}} \rho  \xdust   \Delta_{j} r=0.
\label{eq_E4}
\end{eqnarray}
Note that differences are now computed downwind ($\Delta_j x \equiv x_{j+1} - x_{j}$), as $v$ and $\tau$ have boundary conditions defined at the sonic point ($j=N$). When evaluating variables in (\ref{eq_E3}) and (\ref{eq_E4}), we average across adjacent gridpoints downwind, e.g., $\rho \equiv (\rho_{j}+\rho_{j+1})/2$.

\subsubsection{Boundary Conditions}

We need four boundary conditions to solve the system of finite difference equations (\ref{eq_E1})--(\ref{eq_E4}). At the base of the atmosphere, we set the boundary conditions for gas density and temperature using relations from \S\ref{sec_atmosphere}: 

\begin{eqnarray}
E_{1,0}&=&\rho_0-\rho\sub{vapor}=0 \cr
E_{2,0}&=&T_0 - T\sub{surface}=0. 
\end{eqnarray}
At our outer boundary --- the sonic point --- we require the bulk velocity of gas to equal the local sound speed, and $\tau$ to equal our specified value (\S\ref{sec_hydro}): 

\begin{eqnarray}
E_{3,N}&=&v_{N}- c\sub{s}=0 \cr
E_{4,N}&=&\tau_N-\tau\sub{s}=0.
\end{eqnarray}

At every step of the iteration we determine $r\sub{s}$ by demanding $dv/dr$ to be finite at the sonic point. As the denominator of equation (\ref{eq_E3}) vanishes at $v=c\sub{s}$, we require that   
\begin{equation}
\left[  \frac{2\gamma kT}{\mu r} - \frac{\gamma-1}{\rho v}(\Gamma\sub{col}+\Gamma\sub{lat})
-\frac{GM}{r^2}+ \frac{3GM_{*}r}{a^3} \right]_{r\sub{s}}=0.
\label{eq_rs}
\end{equation}

\subsubsection{Method of Solution}
\label{sec_method}

The relaxation method determines the solution by starting with an
appropriate guess and iteratively improving it. We use the
multidimensional Newton's method as our iteration scheme, which
requires us to evaluate partial derivatives of all $E_{i,j}$ with
respect to all $4N$ dependent variables ($\rho_j, T_j, v_j,
\tau_j$). We evaluate partial derivatives numerically, by introducing
changes of order $10^{-8}$ in dependent variables and computing the
appropriate finite differences. Newton's method produces a $4N \times
4N$ matrix, which we invert using the \texttt{numpy} library in
\texttt{python}. At each iteration we numerically solve equation
(\ref{eq_rs}) for $r\sub{s}$ and re-map all $N$ gridpoints between $R$
and $r\sub{s}$. We iterate until variables change by less than one
part in $10^{10}$.

We found our iteration scheme to be rather fragile. The code only
converges if initial guesses are already close to the solution. For
this reason, we started with a simplified version of the problem with
a known analytic solution as the initial guess. We then gradually
added the missing physics to the code, obtaining a solution with each
new physics input until the full problem was solved. Specifically, we
began by solving the isothermal Parker wind problem (e.g.,
\citealt{Lamers:1999p15622}). We enforced the isothermal condition by
declaring $\gamma$ to be unity, so that our energy equation read
$dT/dr=0$ (see eq. \ref{eq_E2}). Additionally, we set $M_{*}$ and
$\xdust$ to zero to remove the effects of tidal gravity and dust. We
then slowly increased each of the parameters $M_{*}$, $\gamma$, and
$x\sub{dust}$ (in that order) to their nominal values. Any subsequent
parameter change (e.g., $M$) was also performed in small increments.


In summary, our code contains three main input parameters:
$x\sub{dust,\,amp}$ and $x\sub{dust,\,max}$ which prescribe the
dust-to-gas profile $\xdust (r)$, and the planet mass $M$. Our goal is
to explore the dependence of the mass loss rate
\begin{equation}
\dot{M}=\Omega\rho v r^2
\label{eq_mdot}
\end{equation}
on these three parameters.  Here $\Omega$ is the solid angle over
which the wind is launched, measured from the center of the planet; we
set $\Omega=1$ since the high surface
temperatures required to produce a wind are likely to be reached only
near the substellar point.  We independently vary the parameters
$x\sub{dust,\,max}$ and $x\sub{dust,\,amp}$ to find the {\it maximum}
$\dot{M}$ for a given $M$.

\section{RESULTS}
\label{sec_results}

\begin{figure}%
\includegraphics[width=\linewidth]{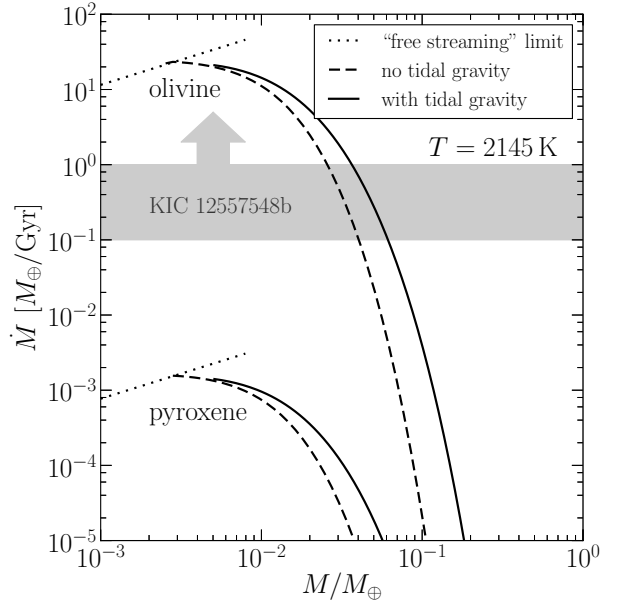}
\caption{%
  Evaporative mass loss rates $\dot{M}$ vs.~planet mass $M$ for
  isothermal \revised{dust-free} winds. Solid lines are for models
  that include stellar tidal gravity, while dashed curves are for
  models that do not. As the planet mass is reduced, all solutions
  converge toward the free-streaming limit where $\dot{M}$ is not
  influenced by gravity but instead scales with the surface area of
  the planet (eq.~\ref{eq_free}). \revised{As inferred from observations, possible present-day mass
    loss rates $\dot{M}\sub{1255b}$ for KIC 1255b are marked in gray. Technically we have only a lower limit on $\dot{M}\sub{1255b}$ of $0.1 M\sub{\earth}$ / Gyr; 
for purposes of discussion throughout this paper, we adopt 0.1--$1 M\sub{\earth}$ / Gyr
as our fiducial range} \revised{(see discussion surrounding equation
    \ref{eq_1255b})}. Clearly KIC 1255b cannot be a pure pyroxene
  planet. Subsequent figures will refer to planets with pure olivine
  surfaces (except in \S\ref{sec_iron} where we consider iron).}
\label{fig_B}
\end{figure}

We begin in \S\ref{sec_isothermal} with an isothermal gas model. The
isothermal model serves both as a limiting case and as a starting
point for developing the full solution which includes a realistic
treatment of the energy equation. We provide results for our full
model in \S\ref{sec_full}. \revised{In the full solution we focus on three
possible planet masses $M=$ 0.01, 0.03, and 0.07 $M\sub{\earth}$,
finding that $M \approx 0.01 \, M\sub{\earth}$ yields a {\it maximum}
$\dot{M}$ that is compatible with the observationally inferred
$\dot{M}\sub{1255b}$. We conclude that in the context of our energy
equation, the present-day mass of KIC 1255b is {\it at most}
$\sim$0.02 $M\sub{\earth}$ (since smaller mass planets can generate
still higher $\dot{M}$ that are also compatible with
$\dot{M}\sub{1255b}$). In \S\ref{sec_histories} we integrate back in
time to compute the {\it maximum} formation mass of KIC 1255b.}

\subsection{Isothermal Solution} \label{sec_isothermal}

We begin by solving a steady, \revised{dust-free}, isothermal wind
described by equations (\ref{eq_mass}) and (\ref{eq_momentum}) with
$M_{*}$ set equal to zero. Although this is a highly idealized
problem, its solution provides us with a starting point (\revised{i.e., an
  initial model}) from which we are able to solve more complicated
problems (see \S\ref{sec_method}). Our code accepts the standard
isothermal wind solution (e.g., Chapter 3 of
\citealt{Lamers:1999p15622}) as an initial guess, with minimal
relaxation.

Mass loss rates derived for the isothermal model ($T$=2145 K, the
temperature given by eq. \ref{eq_Tsurf} with
$\tau\sub{surface}=\tau\sub{s}=0.01$) are plotted in Figure
\ref{fig_B}. They depend strongly on planet mass, with large gains in
$\dot{M}$ for comparatively small reductions in $M$, for the basic
reason that atmospheric densities are exponentially sensitive to
surface gravity. Tidal gravity boosts $\dot{M}$ by reducing the total
effective gravity. As $M$ decreases, gravity becomes increasingly
irrelevant, the thermal speed of the gas eventually exceeds the
surface escape velocity, and the dependence of $\dot{M}$ on $M$
weakens. In the ``free-streaming'' limit (dotted lines in Figure
\ref{fig_B}), mass loss is no longer influenced by gravity, and occurs
at a rate
\begin{equation}
\dot{M}\sub{free} \approx \rho\sub{vapor}c\sub{s}R^2.
\label{eq_free}
\end{equation}
Note that $\dot{M}\sub{free}$ decreases when $M$ is reduced, as less
surface area ($R^2 \propto M^{2/3}$) is available for evaporation. For
$M \sim 0.03 M\sub{\earth}$, a pure olivine surface can produce a
(\revised{dust-free}, isothermal) wind for which $\dot{M}$ becomes
compatible with the observationally inferred value for KIC 1255b, on
the order of $\dot{M}\sub{1255b} \sim 1$ $M\sub{\earth}$/Gyr (this is
the observed mass loss rate for both dust and gas combined;
\revised{see our discussion surrounding equation \ref{eq_1255b}} and
\citealt{Rappaport:2012p14995}).  By contrast, for a pure pyroxene
surface, $\dot{M} \ll \dot{M}\sub{1255b}$ always. Henceforth we will
calculate the density $\rho\sub{vapor}$ of gas at the surface using
parameters appropriate for olivine.

\subsection{Full Solution} \label{sec_full}

\begin{figure}%
\includegraphics[width=\linewidth]{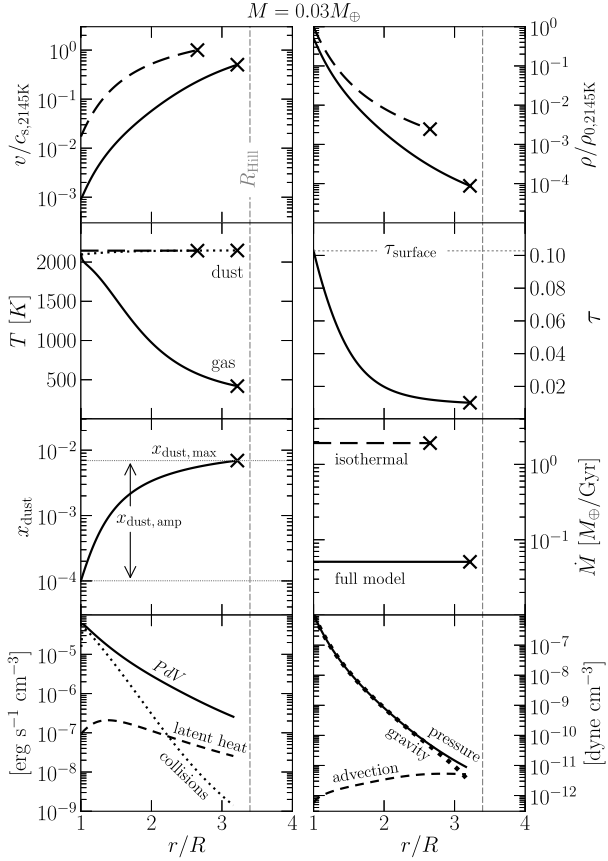}
\caption{%
Radial dependence of wind properties in the full solution for a planet of mass $M=0.03M\sub{\earth}$. Parameters of the \xdust function are chosen to maximize $\dot{M}$. In the upper six panels, the full solution is shown by solid curves, while the \revised{dust-free} isothermal solution is marked by dashed curves. \revised{Values for $v$ have been normalized to the sound speed at $T=2145$ K, and values for $\rho$ have been normalized to the density $\rho\sub{0,2145K}=\mu P\sub{vapor}/(kT)$ evaluated also at $T = 2145$ K.} The $\times$-symbol marks the location of the sonic point (the outer boundary of our solution), which occurs close to the Hill radius $R\sub{Hill}$ (marked by a vertical line). The two lower panels show the contributions of the individual terms in the energy and momentum equations (left and right panels, respectively).}
\label{fig_C-0_03}
\end{figure}

We relax the isothermal condition in our code by slowly increasing $\gamma$ from 1 to 1.3. With this parameter change, we are solving the full energy equation (eq.~\ref{eq_energy}). Grains are gradually added by modifying $x\sub{dust,\,amp}$ and $x\sub{dust,\,max}$, necessitating the calculation of optical depth (eq.~\ref{eq_tau}).

In Figure \ref{fig_C-0_03} we show the solution for which $\dot{M}$
was maximized (over the space of possible values of $\xdust$) for
$M=0.03$ $M\sub{\earth}$, or about half the mass of Mercury. Recall
from \S\ref{sec_isothermal} that when the wind was assumed to be
isothermal, a planet of this mass was able to reproduce the inferred
mass loss rate of KIC 1255b. Now, with our treatment of the full
energy equation and the inclusion of dust, $\max \dot{M}$ plummets by
a factor of 40. Relative to the isothermal solution,
the reduction in mass loss rate in the full model is mainly caused by
the gas temperature dropping from 2095 K at the planet surface 
to below 500 K at the sonic point, and the consequent reduction in gas
pressure. The temperature drops because gas expands in the wind and
does $PdV$ work. Gas heating by dust-gas collisions or grain formation
could, in principle, offset some of the temperature reduction, but
having too many grains also obscures the surface from incident
starlight, decreasing $T\sub{surface}$ and $\rho\sub{vapor}$ (the
latter quantity depends exponentially on the former). The particular
dust-to-gas profile $\xdust$ that is used in Figure \ref{fig_C-0_03}
is such that the ability of dust to heat gas is balanced against the
attenuation of stellar flux by dust, so that $\dot{M}$ is maximized
(for this $M$).

Note that the flow at the base begins with velocities of
$\sim$$10^{-3}$ the sound speed. At these subsonic velocities, the
atmosphere is practically in hydrostatic equilibrium, with gas
pressure and gravity nearly balancing. Only when velocities are nearly
sonic does advection become a significant term in the momentum
equation (bottom right panel of Figure \ref{fig_C-0_03}). 
Flow speeds are still high enough at the base of the
  atmosphere to lift dust grains against gravity for all but the
  highest planet masses considered (\S\ref{sec_1fluid}; gas speeds as
  low as $\sim$1 m/s in a $\sim$$1 \,\mu$bar atmosphere can enable
  micron-sized grains to escape sub-Mercury-sized planets, as can be
  verified by equation \ref{eq_blowout}; furthermore, the condition
  that grains not slip relative to gas is conveniently independent of
  gas velocity in the Epstein free-molecular drag regime, as can be
  seen in equation \ref{eq_stop}. \rev{Our flows are entirely
in the Epstein drag regime, as explained below equation \ref{eq_stop}}.)

\begin{figure*}
\centering
\includegraphics[width=0.49\linewidth]{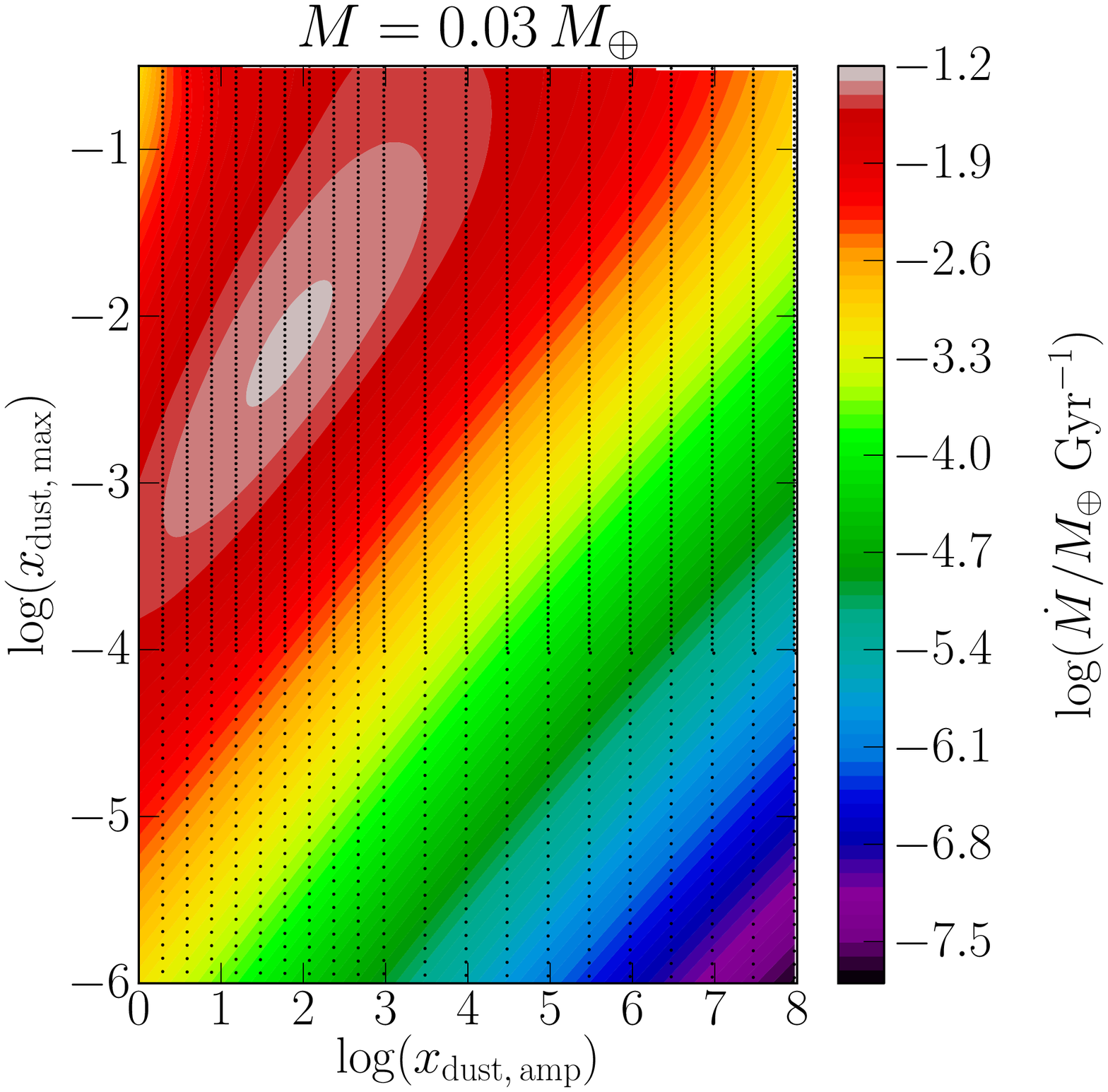}
\includegraphics[width=0.49\linewidth]{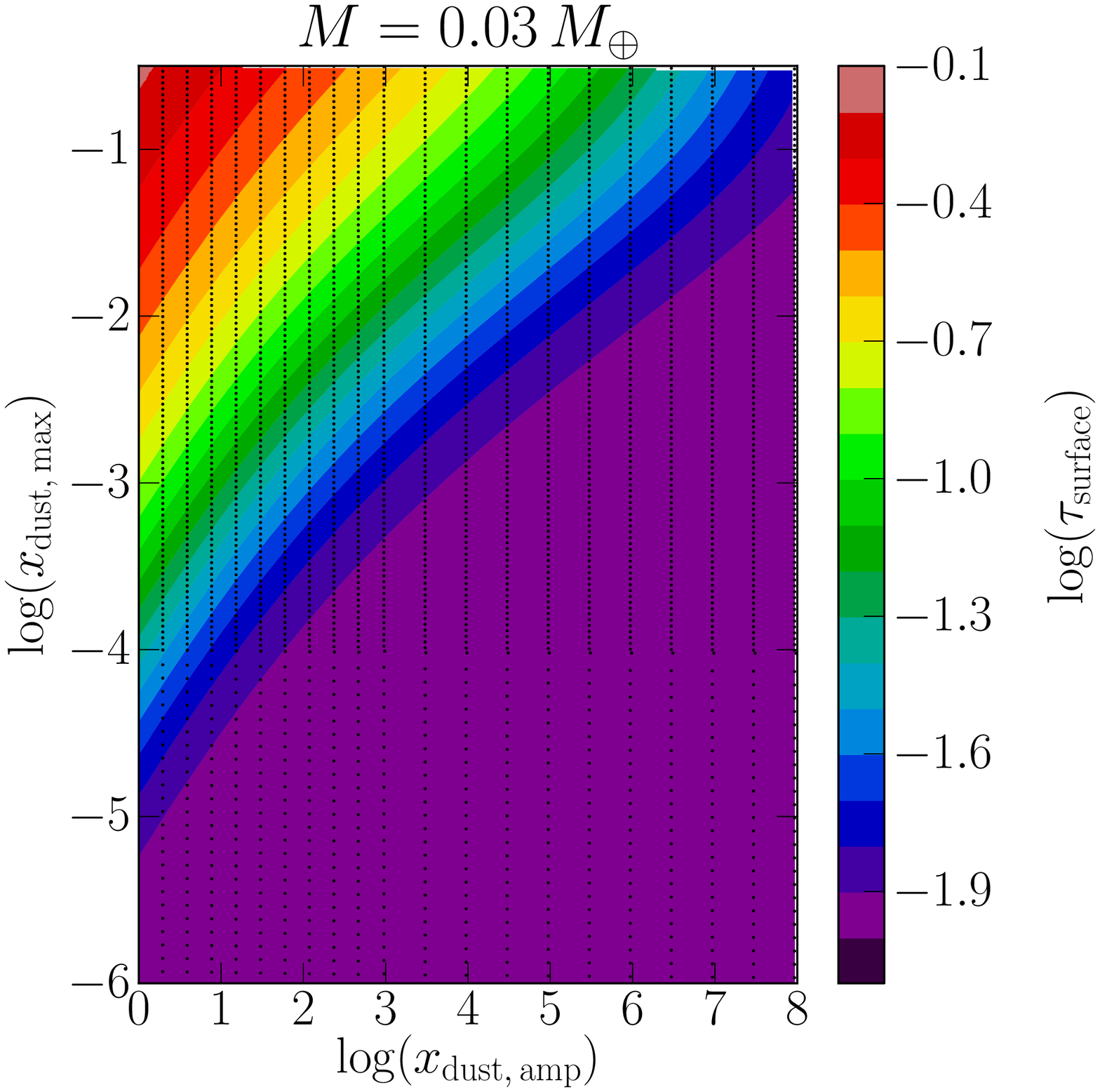}
\caption{%
{\it Left panel}: Dependence of $\dot{M}$ on dust abundance for a planet of mass $M=0.03M\sub{\earth}$. The dust abundance parameters $\xmax$ and $\xamp$ are defined in equation (\ref{eq_xdust}); see also Figure \ref{fig_C-0_03}. Both parameters are varied independently to find the maximum $\dot{M}$, which occurs at log($x\sub{dust,\,amp}$, $x\sub{dust,\,max}$) $\sim$ ($1.7$, $-2.2$). The full model corresponding to this maximum $\dot{M}$ is detailed in Figure \ref{fig_C-0_03}. Each black dot corresponds to a full model solution. Colour contours for $\dot{M}$ are interpolated using a cubic polynomial. {\it Right panel}: Optical depth $\tau\sub{surface}$ between the star and the planetary surface as a function of dust abundance for a planet of mass $M=0.03 M\sub{\earth}$. Apart from the region near the peak in $\dot{M}$, contours of constant $\tau\sub{surface}$ are roughly parallel to contours of constant $\dot{M}$, suggesting that the value of $\tau\sub{surface}$ is more important for determining $\dot{M}$ than the specific functional form of $x\sub{dust}$.}
\label{fig_D-0_03}
\end{figure*}

In Figure \ref{fig_D-0_03}, we show how $\dot{M}$ and
$\tau\sub{surface}$ vary with the function $x\sub{dust}$. As just
discussed, $\max \dot{M}$ is reached for a specific choice of
parameters which balance gas heating by dust and surface obscuration
by dust. \revised{(We emphasize, here and elsewhere, that we
  have not identified a physical reason why actual systems like KIC
  1255b should have mass loss rates $\dot{M}$ that equal their
  theoretically allowed maximum values.)}
Apart from the region
near the peak in $\dot{M}$, contours of constant $\tau\sub{surface}$
are roughly parallel to contours of constant $\dot{M}$, suggesting
that the value of $\tau\sub{surface}$ is more important for
determining $\dot{M}$ than the specific functional form of
$x\sub{dust}$. This finding increases the confidence we have in the
robustness of our solution.

\begin{figure}%
\includegraphics[width=\linewidth]{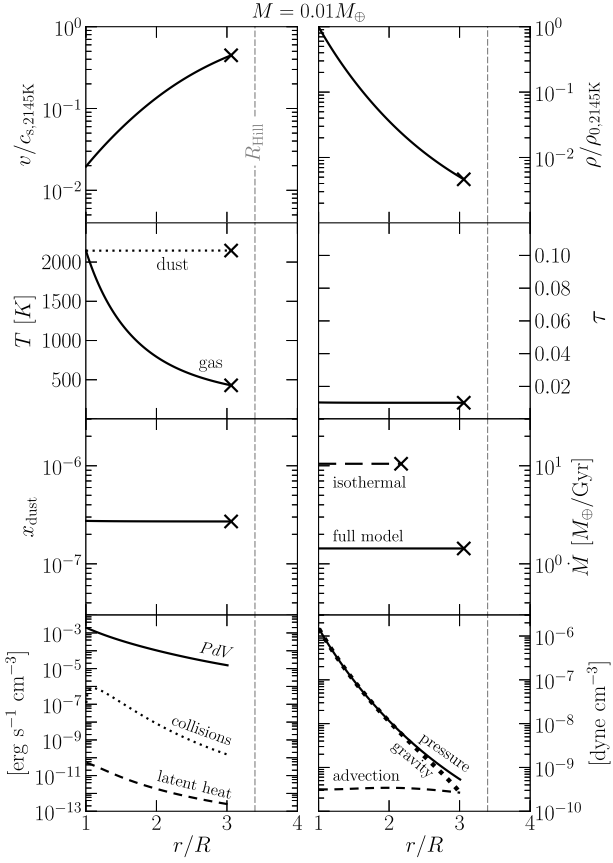}
\caption{%
Same as Figure \ref{fig_C-0_03}, but for a planet of mass $M=0.01 M\sub{\earth}$. For this planet mass, $\dot{M}$ is maximized when essentially no dust is present (i.e., the atmosphere is essentially transparent) as gas moves too quickly for dust-gas collisions to heat the gas. As such, all heating terms are negligible and the gas expands adiabatically.}
\label{fig_C-0_0133}
\end{figure}

\revised{It is clear from Figures \ref{fig_C-0_03} and
  \ref{fig_D-0_03} that a planet of mass $M=0.03$ $M\sub{\earth}$ can
  only emit winds for which $\max \dot{M} < \min \dot{M}\sub{1255b}
  \sim 0.1 M\sub{\earth}$ / Gyr (at least within the context of our
  energy equation). In Figure \ref{fig_C-0_0133} we show results for
  $M=0.01$ $M\sub{\earth}$, for which $\max \dot{M} > \min
  \dot{M}\sub{1255b}$.} For this $M$, the mass loss rate $\dot{M}$ is
maximized when the planet surface is essentially unobscured by
dust. At our arbitrarily chosen value for $\xdust\sim 3\e{-7}$, the
wind is not significantly heated by dust and expands practically
adiabatically. The maximum $\dot{M}$ is reached for an essentially
dust-free solution because the wind is blowing at too high a speed for
dust-gas collisions to be important. That is, the timescale over which
gas travels from the planet's surface to the sonic point is shorter
than the time it takes a gas particle to collide with a dust grain:
gas is thermally decoupled from dust.  Were we to increase \xdust to
make dust-gas heating important, the flow would become optically thick
and the wind would shut down. If the model shown in Figure
\ref{fig_C-0_0133} does represent KIC 1255b, then the dust grains that
occult the star must condense outside the sonic point, beyond the Hill
sphere.

\begin{figure}
\includegraphics[width=\linewidth]{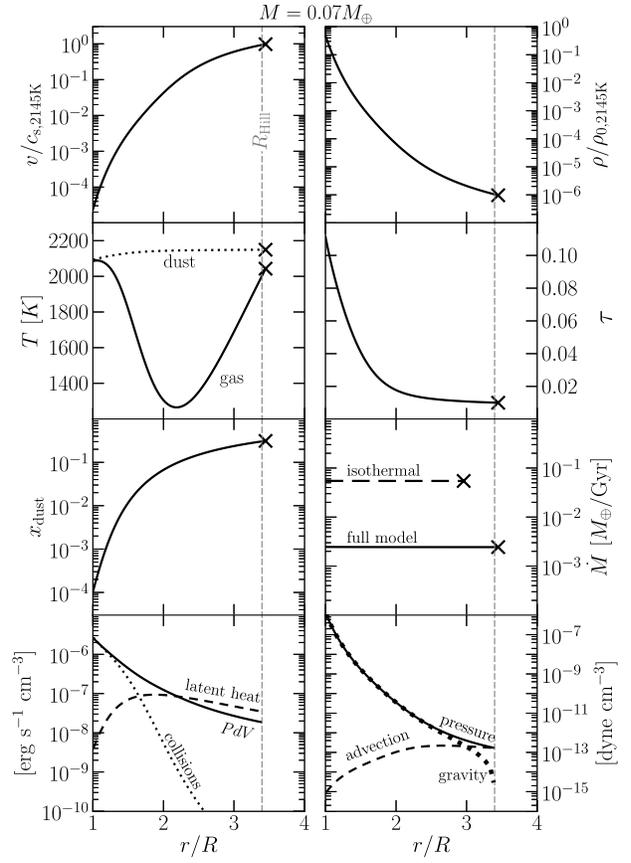}
\caption{%
Same as Figure \ref{fig_C-0_03}, but for a planet of mass $M=0.07 M\sub{\earth}$. For this mass, $\dot{M}$ is maximized for the dustiest flows considered in this paper. The outflow is launched at such low velocities that collisional dust-gas heating is important. Near the sonic point, latent heating overtakes $PdV$ cooling and the temperature of the gas rises.}
\label{fig_C-0_0675}
\end{figure}

Figure \ref{fig_C-0_0675} shows the solution for $M=0.07$ $M\sub{\earth}$. In this case $\dot{M}$ is maximized when enough dust is present to heat the gas significantly above the adiabat. At this comparatively large planet mass, initial wind speeds are low enough that a gas particle collides many times with dust over the gas travel time, so that dust-gas collisions heat the gas effectively near the surface of the planet. %
Further downwind, near the sonic point, latent heating overtakes $PdV$ cooling and actually increases the temperature of the gas with increasing altitude.
The wind is much more sensitive to heating near the base of the flow than near the sonic point: if we omit latent heating from our model --- so that gas cools to $\sim$500 K at the sonic point --- then $\dot{M}$ is reduced by only a factor of two. In comparison, $\dot{M}$ drops by several orders of magnitude if dust-gas collisions are omitted.

\begin{figure*}
\centering
\includegraphics[width=0.49\linewidth]{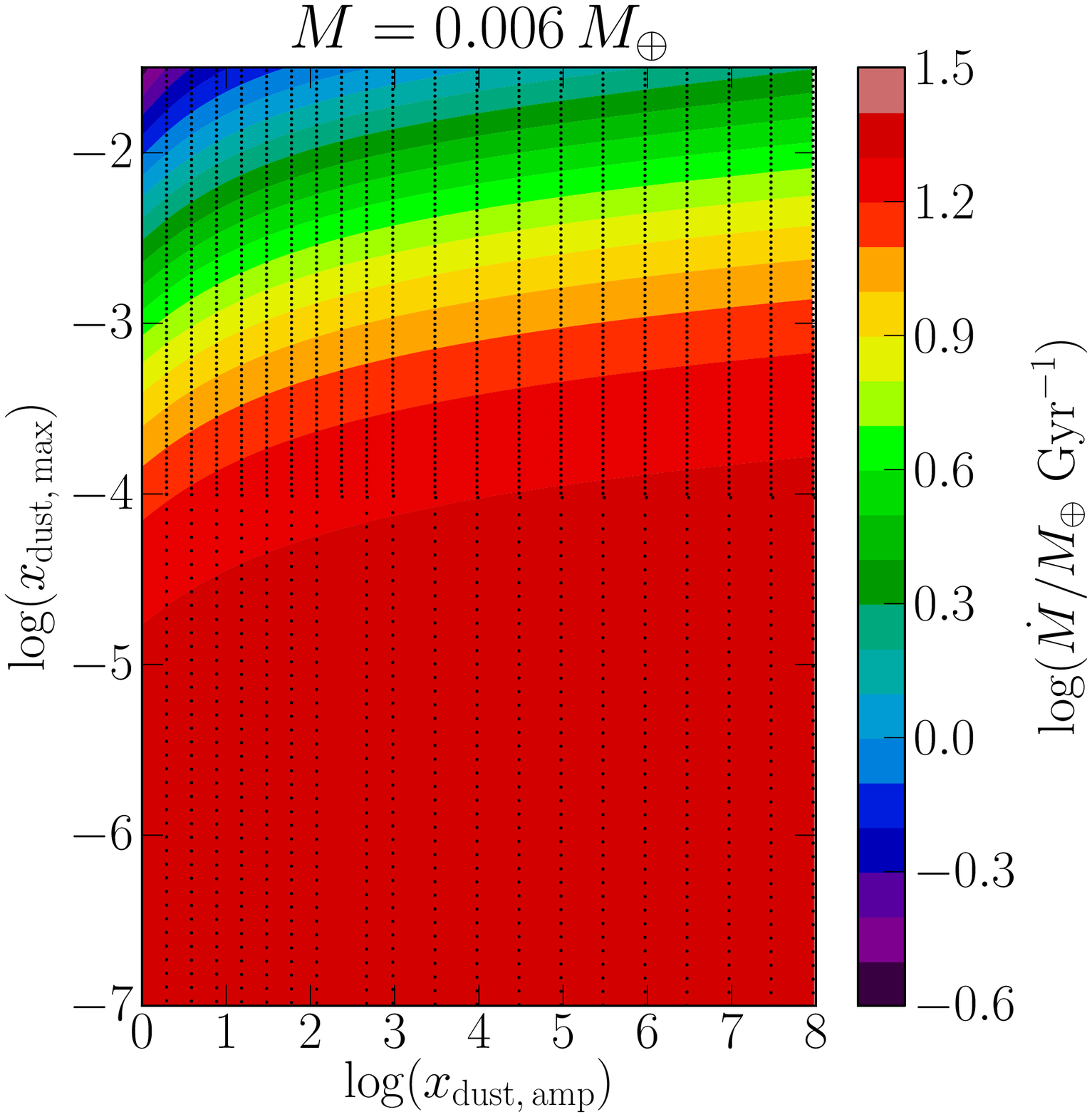}
\includegraphics[width=0.49\linewidth]{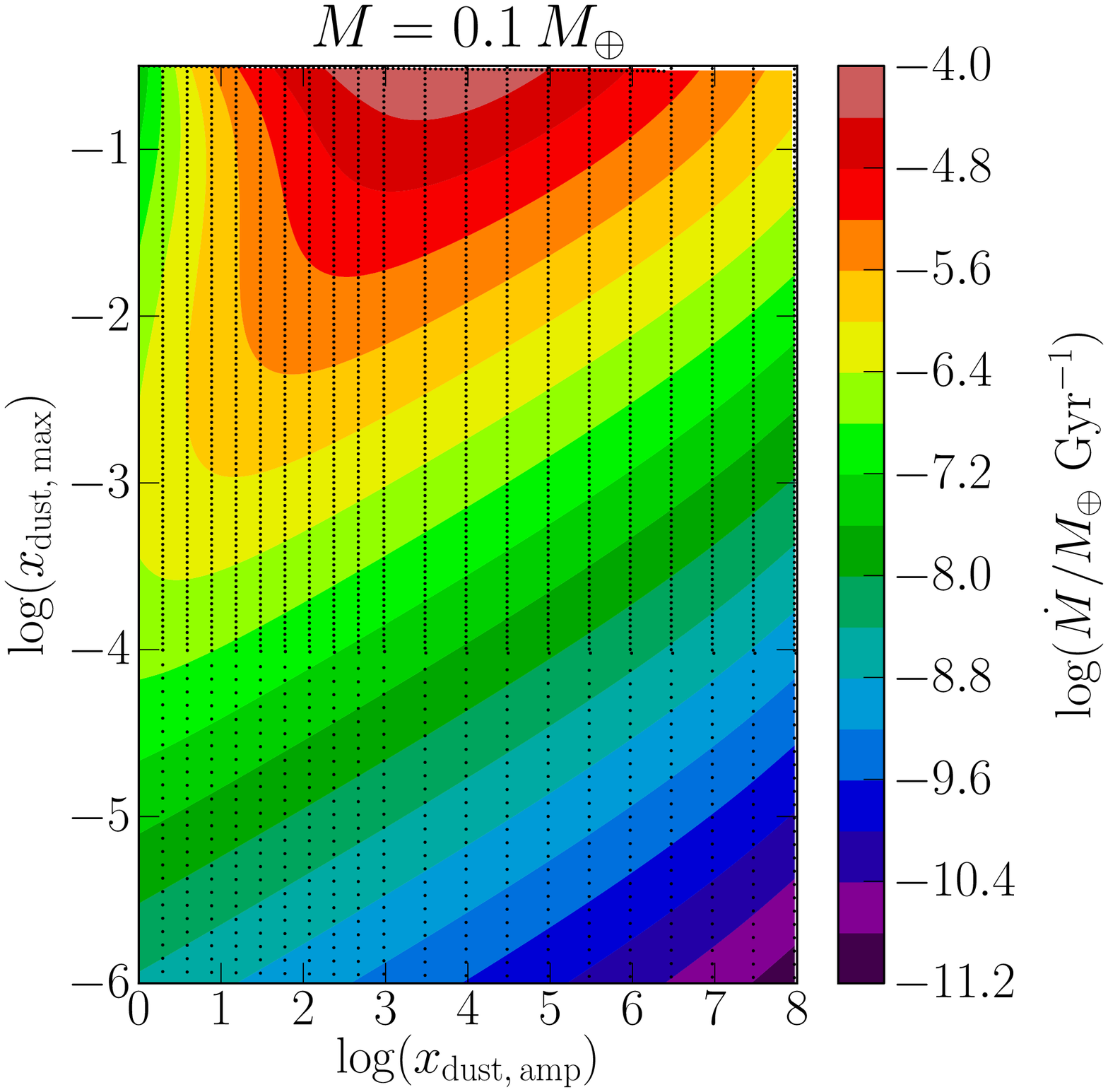}
\caption{%
{\it Left panel}: Same as the left panel of Figure \ref{fig_D-0_03}, but for a planet of mass $M=0.006 M\sub{\earth}$. In this low-mass limit, $\dot{M}$ is maximized when no dust is present (i.e., when the atmosphere is transparent) because gas is moving too quickly for heating by dust to be significant. {\it Right panel}: Same as left panel, but for a planet of mass $M=0.1 M\sub{\earth}$. In this high-mass limit, $\dot{M}$ is maximized for the dustiest flows we consider (i.e., the highest values of $x\sub{dust,max}$) because flow speeds near the surface are slow enough for dust-gas energy exchange to be significant.}
\label{fig_E}
\end{figure*}

The dependence of $\dot{M}$ on \xdust in both the ``dust-free
low-mass'' and the ``dusty high-mass'' limits is illustrated in Figure
\ref{fig_E}. In the low-mass limit, $\dot{M} = \max \dot{M}$ when no
dust is present.  In this regime, gas moves too quickly for heating by
dust to be significant; dust influences the flow only by attenuating
starlight, and as long as $\tau\sub{surface} \ll 1$, dust hardly
affects $\dot{M}$. By contrast, in the high-mass limit, $\dot{M} =
\max\dot{M}$ for the dustiest flows we consider (i.e., the highest
values of \xmax).  In this regime, $\dot{M}$ is very sensitive to 
dust abundance, with values spanning six orders of magnitude over the
explored parameter space.

In Figure \ref{fig_F} we show mass loss rates as a function of planet
mass for the full model. We emphasize that these are maximum mass loss
rates, found by varying $x\sub{dust}$. At low planet masses, $\max
\dot{M}$ for the full model converges with $\dot{M}$ for the
isothermal model because both approach the free-streaming limit, where
gravity becomes irrelevant and $\dot{M}$ is set entirely by conditions
at the surface of the planet (eq.~\ref{eq_free}). \revised{To reach
  $\dot{M}\sub{1255b} > 0.1 M\sub{\earth}$ / Gyr, the present-day mass
  of KIC 1255b must be $\lesssim 0.02 \, M\sub{\earth}$, or less than
  about twice a lunar mass.}

\revised{We have verified {\it a posteriori} for the models shown in Figures \ref{fig_C-0_03}, \ref{fig_C-0_0133}, and \ref{fig_C-0_0675} that
the sonic point is attained at an altitude
where the collisional mean free path of gas molecules
is smaller than $r\sub{s}$, so that the hydrodynamic
approximation embodied in equations (\ref{eq_mass})--(\ref{eq_energy})
is valid. In other words, the exobase lies outside the
sonic point in these $\dot{M}=\max \dot{M}$ models. The margin of safety is largest for the lowest mass models. 
}

\subsection{Mass-Loss Histories} 
\label{sec_histories}

We calculate mass-loss histories $M(t)$ by time-integrating our
solution, shown in Figure \ref{fig_F}, for $\max
\dot{M}(M)$. Before we integrate, however, we introduce $0 <
f\sub{duty} < 1$ to account for the duty cycle of the wind: we define
$f\sub{duty} \cdot \max \dot{M}$ as the time-averaged mass-loss
rate. We estimate that $f\sub{duty} \sim 0.5$ based on the
statistics of transit depths compiled by Brogi et
al.~(\citeyear{Brogi:2012p15046}, see their figure 2).  We
time-integrate $f\sub{duty} \cdot \max \dot{M}$ to obtain the $M(t)$
curves shown in Figure \ref{fig_G}. (This factor of 2 correction
for the duty cycle should not obscure the fact
that the actual mass loss rate, time-averaged or otherwise, could still
be much lower than the theoretically allowed $\max \dot{M}$,
a possibility we return to throughout this paper.)

We highlight the case of a planet with a lifetime of $t\sub{life} = 5$
Gyr. Under our full (non-isothermal) model (right panel of Figure
\ref{fig_G}), such a planet has a mass at formation of 0.06
$M\sub{\earth}$ and gradually erodes over several Gyr until it reaches
$M \sim 0.03$ $M\sub{\earth}$ --- whereupon the remaining mass is lost
catastrophically over a short time (just how short is estimated for
the specific case of KIC 1255b below).  Contrast this example with
planets having initial masses $\gtrsim 0.07$ $M\sub{\earth}$ --- these
have such low $\dot{M}$ that they barely lose any mass over Gyr
timescales.

For comparison, we also present mass-loss histories using a dust-free
isothermal model with stellar tidal gravity (left panel of Figure
\ref{fig_G}).  Compared to our non-isothermal solutions --- for which
gas temperatures fall immediately as the wind lifts off the planet
surface --- the isothermal solution corresponds to a flow which stays
relatively pressurized and which therefore enjoys the largest
$\dot{M}$ for a given $M$.  The isothermal wind thus represents an
endmember case. However, the mass-loss histories under the isothermal
approximation do not differ qualitatively from those using the full
energy equation. For isothermal winds, the dividing mass between
planet survival and destruction within 10 Gyr is about $0.11
\, M\sub{\earth}$, only $\sim$40\% larger than the value cited above for
our non-isothermal solutions.

\subsection{\revised{Possible Mass Loss Histories for KIC 1255b}} \label{sec_apply}

\begin{figure}%
\includegraphics[width=\linewidth]{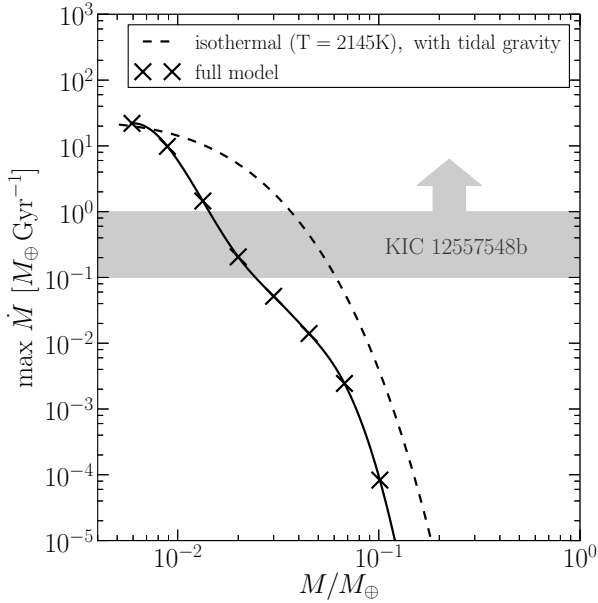}
\caption{%
  Maximum mass loss rates $\dot{M}$ vs.~planet mass $M$ for the full
  model. At each $\times$-marked mass, max $\dot{M}$ was found by
  varying $\xdust$ as described, e.g., in Figure \ref{fig_D-0_03}. The
  solid curve is a cubic spline interpolation. Mass loss rates for the
  full model are generally lower than for the isothermal model (dashed
  curve) because the dust-gas heating terms we have included in our
  full model turn out to be inefficient. At low planet masses, the
  isothermal and full models converge because both approach the
  free-streaming limit, where gravity becomes irrelevant and $\dot{M}$
  is set entirely by conditions at the surface (i.e., surface area,
  equilibrium vapor pressure, and sound speed;
  eq. \ref{eq_free}). \revised{According to the full model, the
  present-day mass of KIC 1255b is $\lesssim 0.02 \, M\sub{\earth}$,
  or less than twice the mass of the Moon; for such masses, $\max \dot{M} > \min \dot{M}_{1255b} \sim 0.1 \,M\sub{\earth}$ / Gyr.}}
\label{fig_F}
\end{figure}

\begin{figure}%
\includegraphics[width=\linewidth]{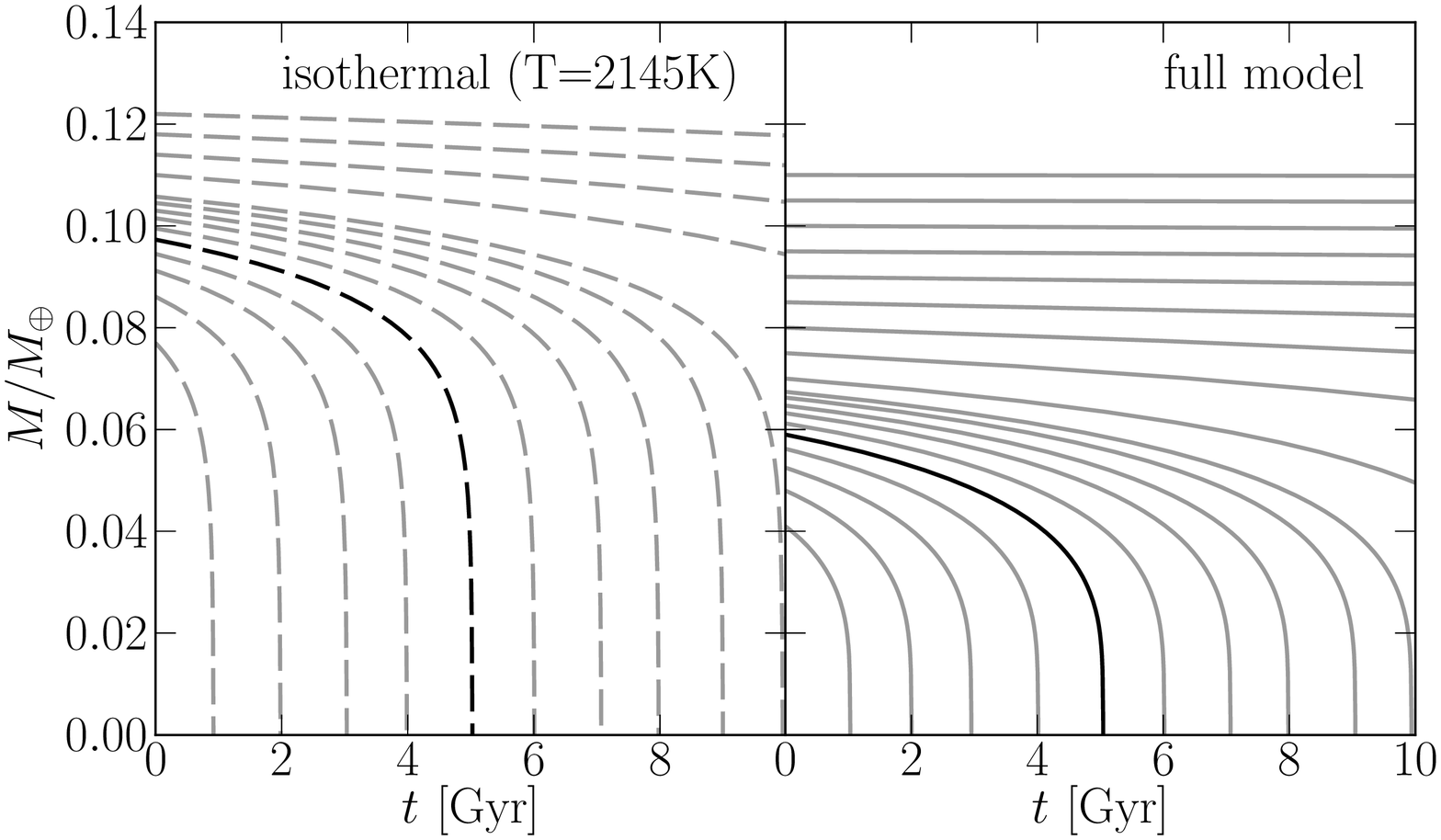}
\caption{ Mass-loss histories $M(t)$ obtained by
  \revised{time-integrating $f\sub{duty} \cdot \max \dot{M}(M)$} with
  $f\sub{duty} = 0.5$ for the isothermal and full models. We highlight
  the case of a planet with a 5 Gyr lifetime. For the full model
  (right panel), this corresponds to an initial mass of $\sim$0.06
  $M\sub{\earth}$, slightly larger than the mass of Mercury. Such a
  planet could have formed \textit{in situ} and slowly eroded over
  several Gyr until reaching $\sim$0.03 $M\sub{\earth}$, whereupon the
  planet evaporates completely in a few hundred Myr. By contrast,
  planets with formation masses $\gtrsim 0.07 M\sub{\earth}$ survive
  (in the full model) for tens of Gyrs without significant mass loss.}
\label{fig_G}
\end{figure}

Figures \ref{fig_F} and \ref{fig_G} can be used to sketch possible
mass-loss histories for KIC 1255b. For our first scenario, we employ
the full model that solves the full energy equation.  \revised{Today,
  to satisfy the observational constraint that $\dot{M}\sub{1255b} > 0.1
  \, M\sub{\earth}/{\rm Gyr}$ (see the derivation of our equation
  \ref{eq_1255b}), the planet must have a present-day mass of {\it at most}
  $M \approx 0.02 \,M\sub\earth$, or roughly twice the mass of the
  Moon (Figure \ref{fig_F}).} Such a low mass implies that currently
KIC 1255b is in a ``catastrophic evaporation'' phase (vertical
straight lines in Figure \ref{fig_G}). Depending on the age of the
planet (i.e., the age of the star: 1--10 Gyr), KIC 1255b
originally had a mass of {\it at most} 0.04--0.07 $M\sub\earth$: 2--4 times its
maximum current mass, or about the mass of Mercury.  
\revised{Starting from
today, the time the planet has before it disintegrates completely is
on the order of $t\sub{evap} \sim M/(f\sub{duty} \cdot \max \dot{M})$,
which is as long as 400 (40) Myr for $M = 0.02 \,M\sub{\earth}$,
$f\sub{duty} = 0.5$, and $\max \dot{M} = 0.1$ (1) $M\sub{\earth}$ /
Gyr.}

If our energy equation were somehow in error and the wind better
described as isothermal, then the numbers cited above would change
somewhat. The present-day mass of KIC 1255b would be at most $M \sim
0.07 M\sub{\earth}$; the maximum formation mass would be between 0.08--0.11
$M\sub{\earth}$; and the evaporation timescale starting from today
could be as long as $t\sub{evap} \sim 400$ Myr if the present-day mass
$M \sim 0.07 M\sub{\earth}$.

\revised{As we have stressed throughout, a critical assumption we have
  made is that the time-averaged mass loss rate is $f\sub{duty} \cdot
  \max \dot{M}$ with $f\sub{duty}$ as high as 0.5. We have not
  identified a physical reason why the actual mass loss rate should be
  comparable to the theoretically allowed $\max \dot{M}$ for a given
  $M$.  If our assumption were in error and actual mass loss rates
  $\dot{M} \ll \max \dot{M}$, then the present-day mass of KIC 1255b,
  the corresponding evaporation time, and the formation mass of KIC
  1255 would all decrease from the upper bounds cited above.}

\revised{The degeneracy of possible masses and evaporative histories
  outlined in this section could be broken with 
  observational searches for the progenitors of KIC 1255b-like objects
  (\S\ref{sec_occurrence}).}

\section{DISCUSSION}
\label{sec_discussion}

We discuss how mass loss is not ``energy-limited'' in
\S\ref{sec_e_limited}; how gas entrains dust in \S\ref{sec_1fluid};
how the evaporative wind can be time-variable in \S\ref{sec_time}; the
effects of winds on orbital evolution in \S\ref{sec_longterm}; the
interaction of the planetary wind with the stellar wind in
\S\ref{sec_flowconfinement}; the occurrence rate of \revised{quiescent}
progenitors of catastrophically evaporating planets in
\S\ref{sec_occurrence}; the possibility that evaporating planets can
be naked iron cores in \S\ref{sec_iron}; and the ability of dust to
condense out of the wind in \S\ref{sec_condensation}.

\subsection{Mass Loss is Not Energy-Limited}
\label{sec_e_limited}

Atmospheric mass loss rates are commonly assumed to be ``UV-energy-limited'': it is assumed that a fixed, order-unity fraction $\epsilon$ of the incident UV flux $F_{\rm UV}$ does $PdV$ work and lifts material out of the planet's gravitational well. Under the energy-limited assumption, the mass loss rate is $\dot{M} \sim \epsilon \pi R^2 F\sub{UV}/(GM/R)$. The validity of this formula has been tested by \citet{Watson:1981p16957} for terrestrial atmospheres and by \citet{MurrayClay:2009p16619} for hot Jupiters.

The energy-limited formula does not apply at all to the evaporating rocky planets considered in this paper. The stellar UV flux is not essential for low-mass planets because their escape velocities are so small that energy deposition by photoionization is not necessary for driving a wind. Even optical photons --- essentially the stellar bolometric spectrum --- can vaporize close-in planets. Merely replacing $F\sub{UV}$ with $F\sub{bolometric}$ in the energy-limited formula would still be misleading, however, because most of the incident energy is radiated away and does no mechanical work. More to the point, $\epsilon$ is not constant; as discussed in \S\ref{sec_full}, the energetics of the flow changes qualitatively with planet mass. 
If we were to insist on using the energy-limited formula, we would find $\epsilon \sim 10^{-8}$ for $M \sim 0.1 M\sub{\earth}$, and $\epsilon \sim 10^{-4}$ for $M \sim 0.01 M\sub{\earth}$ (see Figure \ref{fig_F}).

\subsection{Dust-Gas Dynamics}
\label{sec_1fluid}

In our model we have assumed that dust grains condensing within the
wind are carried along without any relative motion between dust and
gas (1-fluid approximation).  Grains are well entrained if their
momentum stopping times
\begin{equation}
t\sub{stop} \sim \frac{m\sub{grain} v\sub{rel}}{\rho v\sub{rel} c\sub{s} s^2} \sim \frac{s \rho\sub{int}}{\rho  c\sub{s}}
\label{eq_stop}
\end{equation}
are shorter than the grain advection time $t\sub{adv} \sim R / v$.  In
equation (\ref{eq_stop}), $m\sub{grain}$ is the mass of an individual
grain and $v\sub{rel}$ is the relative gas-grain velocity. The
aerodynamic drag force in the denominator is given by the Epstein law,
which is appropriate for $v\sub{rel} \lesssim c\sub{s}$ and grain
sizes $s$ smaller than the gas collisional mean free path (for our
flows, $\lambda\sub{mfp} \sim \mu/(\rho \sigma) \gtrsim 2$ cm, where
$\sigma \sim 3\e{-15}$ cm$^2$).

We find using our full model (for $\dot{M} = \max \dot{M}$) that
$t\sub{stop}/t\sub{adv} < 1$ everywhere for $M < 0.03 \,
M\sub{\earth}$, confirming our assumption that micron-sized (and
smaller) grains are well-entrained in the winds emanating from such
low-mass planets.  When $M = 0.03 \,M_{\earth}$, $t\sub{stop} <
t\sub{adv}$ close to the planet's surface, but as gas nears the sonic
point, $t\sub{stop}$ becomes comparable to $t\sub{adv}$. For $M > 0.03\,
M_{\earth}$, the 1-fluid approximation breaks down near the sonic
point. For such large planet masses, future models should account for
gas-grain relative motion --- in addition to other effects that become
increasingly important near the sonic point / Hill sphere boundary
(e.g., Coriolis forces and stellar radiation pressure).  Note that the
1-fluid approximation should be valid for the present-day dynamics of
KIC 1255b, since according to the full model its current mass is at
most $M \approx 0.02 \,M\sub{\earth} < 0.03 \,M\sub{\earth}$.

Although dust can slip relative to gas, dust can still be transported
outward and escape the planet if the aerodynamic drag force exceeds
the force of gravity:
\begin{equation}
	\rho v\sub{rel} c\sub{s} s^2 \gtrsim  \rho\sub{int}s^3 g,
	\label{eq_blowout}
\end{equation}
where $g=G(M/r^2-3M_{*}r/a^3)$ is the total gravitational
acceleration.  Grains that manage to be lifted beyond the Hill sphere (which is situated
close to the sonic point in all our models) are no longer bound to the
planet. Outside the Hill sphere, dust decouples from gas
and is swept into a comet-like tail by stellar radiation pressure and Coriolis
forces.

For $M \approx 0.03 \,M_{\earth}$ and our adopted grain size $s = 1\,
\mu$m, the forces in equation (\ref{eq_blowout}) are comparable near
the sonic point when $v\sub{rel} \sim c\sub{s}$ (i.e., when grains are barely
lifted by drag).
For $M > 0.03 \,M\sub{\earth}$, micron-sized and larger grains will
not be dragged past the sonic point according to (\ref{eq_blowout})
--- thus our assumption that they do fails for such high-mass planets,
and will need to be rectified in future models. Smaller,
sub-micron-sized grains can, however, be lifted outward. Moreover,
sub-micron sized grains may heat the gas more effectively because they
have greater geometric surface area for dust-gas collisions (at fixed
$\xdust$) and because their temperatures exceed those of blackbodies
(their efficiencies for emission at infrared wavelengths are much less
than their absorption efficiencies at optical wavelengths).
The superior momentum and thermal coupling enjoyed
by grains having sizes $s \ll 1 \,\mu$m, plus their relative
transparency at optical wavelengths --- which helps them avoid
shadowing the planet surface from stellar radiation --- motivate
their inclusion in the next generation of models.

Our model could be improved still further by self-consistently
accounting for how the gas density $\rho$ must decrease as dust grains
condense out of gas.  Moreover, the restriction that $\xdust < 1$
could be lifted.

\subsection{Time Variability}
\label{sec_time}

Occultations of KIC 12557548 vary in depth from a maximum of 1.3\% to
a minimum of $\lesssim 0.2$\% on orbital timescales.
\citet{Rappaport:2012p14995} discussed qualitatively the possibility
that such transit depth variations arise from a limit cycle that
alternates between high-$\dot{M}$ and low-$\dot{M}$ phases. A
high-$\dot{M}$ phase that produces a deep eclipse would also shadow
the planet surface from starlight. The resultant cooling would lower
the surface vapor pressure and lead to a low-$\dot{M}$ phase --- after
which the atmosphere would clear, the surface would re-heat, and the
cycle would begin anew.  Such limit cycle behavior could be punctuated
by random explosive events that release dust, similar to those
observed on Io (\citealt{Geissler:2003p16007}; see also the references
in \S4.2 of \citealt{Rappaport:2012p14995}).

Order-of-magnitude variations in $\dot{M}$ arise from only small
fractional changes in surface temperature because the vapor pressure
of gas over rock depends exponentially on temperature. For example,
from equations (\ref{eq_Tsurf}) and (\ref{eq_pvap}), we see that
increasing the surface optical depth $\tau\sub{surface}$ from 0.1 to
0.4 reduces the surface temperature $T\sub{surface}$ by 150 K and the
base density $\rho\sub{vapor}$ by a factor of 10. \revised{
Such changes would cause $\dot{M}$ to drop by more than a factor of 10,
because $\dot{M}$ scales super-linearly with $\rho\sub{vapor}$: a linear
dependence results simply because $\dot{M}$ scales with gas density,
while an additional dependence arises because the wind speed
increases with the gas pressure gradient, which in turn scales as
$\rho\sub{vapor} T\sub{surface}$.
}

To reproduce the orbit-to-orbit variations in transit depth observed
for KIC 1255b, the dynamical time $t\sub{dyn}$ of the wind cannot be much longer than the planet's orbital period of $P\sub{orb} = 15.7$
hr.  The dynamical time is that required for dust to be advected from
the planet surface to the end of the comet-like tail that occults the
star: it is the minimum timescale over which the planet's transit
signature ``refreshes''. If $t\sub{dyn} \gg P\sub{orb}$, then we would
expect transit depths to be correlated from one orbit to the next --- in
violation of the observations.

Referring to the full model for a KIC 1255b-like mass of $M \approx
0.01 M\sub{\earth}$ (Figure \ref{fig_C-0_0133}), we estimate that
\begin{equation}     
t\sub{dyn} =\int_R^{r\sub{s}} \frac{dr}{v} + \int_{r\sub{s}}^{0.1 R_\star} \frac{dr}{v} \sim 13 \, {\rm hr} + \frac{0.1R_\ast}{v\sub{2}} \sim 14\,\,\rm{hr},
\label{eq_tdyn}
\end{equation}
where we have split $t\sub{dyn}$ into two parts: the first integral is
the time for dust to reach the sonic point (the outer boundary of our
calculation), while the second integral is the time for dust to travel
out to $0.1 R_\ast$ (a circular, optically disk of this radius would
generate a 1\% transit depth). The first integral is performed
numerically using our full model (upper left panel of Figure
\ref{fig_C-0_0133}), while the second integral is estimated to
order-of-magnitude using a characteristic grain velocity (well outside
the planet's Hill sphere) of $v\sub{2} \sim 10$ km s$^{-1}$
(\citealt{Rappaport:2012p14995}; see their equation 6 and related
commentary). Since $t\sub{dyn} \lesssim P_{\rm orb}$, we conclude that
the wind/cometary tail can refresh itself quickly enough to change its
appearance from orbit to orbit.

We can go one step further. The timescale over which $\dot{M}$ changes
should be the timescale over which the stellar insolation at the
planetary surface changes --- in other words, the timescale over which
the ``weather'' at the substellar point, where the wind is launched,
changes from, e.g., ``overcast'' to ``clear''. Ignoring the
possibility of volcanic eruptions, we estimate this variability
timescale as the time for the wind to reach the Hill sphere boundary,
at which point the Coriolis force has turned the wind by an
order-unity angle away from the substellar ray joining the planet to
the star (i.e., beyond the Hill sphere, the dust-laden
wind no longer blocks stellar radiation from hitting the substellar
region where the wind is launched).  Because the Hill radius is
situated near the sonic point, this variability timescale is given
approximately by the first integral in equation (\ref{eq_tdyn}):
$t\sub{dyn}^a \approx 13$ hr.

The fact that $t\sub{dyn}^a$ is neither much shorter than nor much
longer than $P\sub{orb}$ supports our proposal that the observed time
variability of KIC 1255b is driven by a kind of limit cycle involving
stellar insolation and mass loss.  Clearly if $t\sub{dyn}^a \gg
P\sub{orb}$, orbit-to-orbit variations in $\dot{M}$ would be
impossible.  Conversely, if $t\sub{dyn}^a \ll P\sub{orb}$ --- or more
precisely if $t\sub{dyn}^a$ were much shorter than the transit
duration of 1.5 hr --- then the wind would vary so rapidly that each
transit observation would time-integrate over many cycles, yielding a
smeared-out average transit depth that would not vary from orbit to
orbit as is observed.

\subsection{Long-Term Orbital Evolution}
\label{sec_longterm}

When computing planet lifetimes in \S\ref{sec_histories}, we have assumed that the planet does not undergo any orbital evolution while losing mass. This approximation is valid because gas leaves the planet's surface at velocities $v\sub{launch} \lesssim c\sub{s} \sim 1$ km s$^{-1}$. Launch velocities are so low compared to the orbital velocity of the planet $v\sub{orb} \sim 200$ km s$^{-1}$ that the total momentum imparted by the wind is a tiny fraction of the planet's orbital momentum.

We quantify this as follows. We estimate the total change in semimajor axis $a$ and eccentricity $e$ from Gauss' equations (see, e.g., \citealt{Murray:1999p16179}). Consider the case in which the gas is launched at a velocity $v\sub{launch}$ at an angle $\alpha$ from the substellar ray and in the plane of the planet's orbit. The angle $\alpha$ could be non-zero because of surface inhomogeneities or asynchronous rotation of the planet. Neglecting terms of order $e^2$ and $\alpha e$, we find that $a$ and $e$ evolve according to 
\begin{eqnarray}
\frac{da}{dt} &\sim & a\frac{v\sub{launch}}{v\sub{orb}} \frac{\dot{M}}{M} \left(e +\alpha\right), \cr
& & \cr
\frac{de}{dt} & \sim  &\frac{v\sub{launch} }{v\sub{orb}} \frac{\dot{M}}{M} (1+\alpha),
\label{eq_gauss}
\end{eqnarray}
where the first term in parentheses on the right-hand-side of each equation is due to the radial component of the perturbation and the second term is due to the azimuthal component (where radius and azimuth are measured in a cylindrical coordinate system centered at the star and in the plane of the planet's orbit). Integrating equations (\ref{eq_gauss}) yields the changes in $a$ and $e$ accumulated over the planet's age: 

\vspace{0.2in}
\begin{eqnarray}
\frac{\Delta a}{a} &\sim & \ln \left( \frac{M\sub{initial}}{M\sub{final}}\right) \frac{v\sub{launch}}{v\sub{orb}} \left(e +\alpha\right) \lesssim 0.01 \left(e +\alpha\right), \cr
& & \cr
\Delta e &\sim & \ln \left( \frac{M\sub{intial}}{M\sub{final}}\right) \frac{v\sub{launch}}{v\sub{orb}} \lesssim  0.01. 
\label{eq_gauss2}
\end{eqnarray}
When evaluating equation (\ref{eq_gauss2}), we have used the maximum possible launch velocity $v\sub{launch} \sim c\sub{s}$ (valid in the ``free-streaming'' limit) and $M\sub{initial}/M\sub{final} \sim 5$ (see \S\ref{sec_histories}).

The fact that evaporating planets undergo negligible orbital evolution
suggests they could have resided on their current orbits for Gyrs ---
indeed they might even have formed \textit{in
  situ}. \citet{Swift:2012p17065} would disfavor \textit{in-situ}
formation of KIC 1225b because at the planet's orbital distance, dust
grains readily sublimate, and ostensibly there would have been no
solid material in the primordial disk out of which rocky planets could
have formed. However, a less strict {\it in-situ} formation scenario
is still viable. Solid bodies larger than dust grains obviously have
longer evaporation times. Such large planetesimals could have drifted
inward through the primordial gas disk and then assembled into the
progenitors of objects like KIC 1255b at their current close-in
distances (e.g., \citealt{Youdin:2002p10876, Hansen:2012p17071, Chiang:2012p16585}). Solid particles can avoid vaporization if they merge
faster than they evaporate.

\subsection{Flow Confinement by Stellar Wind}
\label{sec_flowconfinement}

When computing the transonic solution for the planetary wind, we
implicitly assumed that the outflow was expanding into a vacuum. In
reality, the outflow from the planet will collide with the stellar
wind and form a ``bubble'' around the planet.\footnote{\rev{The considerations in this section parallel those for hot Jupiter winds and their bow shocks;
see \citet{Tremblin:2013p16188}, \citet{Vidotto:2010p17160, Vidotto:2011p17172,
      Vidotto:2011p17167}, and \citet{Llama:2011p17173}.}}
For our solution to be
valid, the radius of the bubble --- i.e., the surface of pressure
balance between the winds --- has to be downstream of the sonic point
where the flow is supersonic. This way, any pressure disturbance
created at the interface cannot propagate upstream and influence our
solution.

At the wind-wind interface, the normal components of the pressures of
the planetary and stellar winds balance. The total pressure includes
the ram, thermal, and magnetic contributions. We compare the pressure
of the stellar wind at $a \sim 0.01 \,\,\rm{AU} \sim 4 R_{*}$ with
that of the planetary wind computed at $r\sub{s}$ to gauge whether the
interface occurs downstream of $r\sub{s}$. For the pressure $P_{*}$ of
a main-sequence, solar-type star we are guided by the Solar wind. Flow
speeds of the Solar wind are measured by tracking the trajectories of
coronal features \citep{Sheeley:1997p16210, Quemerais:2007p16209}. At
$a=0.01$ AU the Solar wind\footnote{These stellar wind parameters are
  appropriate for the ``slow'' Solar wind blowing primarily near the
  equatorial plane of the Sun. In addition, the Solar wind also
  contains a ``fast'' component which emerges from coronal
  holes. During Solar minimum, the fast Solar wind is confined mainly
  to large heliographic latitudes, but may reach the equator plane
  during periods of increased solar activity
  \citep{Kohl:1998p16237,McComas:2003p16247}. Because most planets are
  expected to orbit near their stellar equatorial planes, we take the
  slow component of the Solar wind to be a better guide.} is still
accelerating with typical flow speeds of $v_{*} \sim 100$ km
s$^{-1}$. For this $v_{*}$, and a canonical Solar mass-loss rate of
$2\e{-14}$ M$\sub{\odot}$ yr$^{-1}$, the proton number density is
$n_{*}= 2\e{5}$ cm$^{-3}$. We take the proton temperature to be $T_{*}
\sim 10^6$ K \citep{Sheeley:1997p16210} and the heliospheric magnetic
field strength to be $B_{*} \approx 0.1$ G at 0.01 AU
\citep{Kim:2012p16512}. The total stellar pressure at 0.01 AU is then
$P_{*} \sim n_{*} m\sub{H} v_{*}^2 + n_{*}kT_{*}+B_{*}^{2}/(8 \pi)
\sim 4\e{-4}$ dyne cm$^{-2}$, with the magnetic pressure dominating
other terms by an order of magnitude.

The ram and thermal pressures of the planet's wind are equal at the
sonic point and add up to \revised{(for $\dot{M} = \max \dot{M}$)}
$P\sub{s} \sim 2\e{-3}$ ($1\e{-4}$) dyne cm$^{-2}$ when $M=0.03
\,\,(0.07)\,\, M\sub{\earth}$. Thus for $M \lesssim 0.03
\,M\sub{\earth}$ \revised{(a range that includes possible present-day masses of
KIC 1255b)}, $P\sub{s} > P_{*}$ so that the planetary wind blows a
bubble that extends beyond the sonic point and the transonic solutions
we have computed are self-consistent. By contrast for $M \sim 0.07
\,M\sub{\earth}$, $P\sub{s} \sim P_{*}/4 < P_{*}$ so that the stellar
wind pressure will balance the planetary wind pressure inside of
$r\sub{s}$. The stellar wind will prevent the planetary wind from
reaching supersonic velocities; the planetary outflow will conform to
a ``breeze'' solution with a reduced $\dot{M}$. If the colliding winds
in our problem behave like those of hot Jupiters and their host stars,
then $P\sub{s} \sim P_{*}/4$ will reduce $\dot{M}$ by $\sim$$30\%$ (see
figure 12 of \citealt{MurrayClay:2009p16619}).

We have found $P_{*}$ to be dominated by magnetic pressure. Since magnetic fields only interact with the ionized component of the planetary wind, and since the planetary wind may not be fully ionized, we might be over-estimating the effect of $P_{*}$. Dust grains in the planetary wind may absorb free charges so that the wind may be too weakly ionized to couple with the stellar magnetic field. %
If we ignore the magnetic contribution to the stellar wind, then $P\sub{s} > P_{*}$ for all planet masses that we have considered, and all our transonic solutions would be self-consistent.

What about the planet's magnetic field? A sufficiently ionized outflow
could be confined by the planet's own magnetic field. To assess the
plausibility of this scenario, we use Mercury's field as a guide for
KIC 1255b. The flyby of \textit{Mariner 10} and recent measurements by
\textit{MESSENGER} show that Mercury possesses a weak magnetic field
with a surface strength of $\sim 3\e{-3}$ Gauss
\citep{Ness:1974p16202, Anderson:2008p16198}. The associated magnetic
pressure $\sim 4\e{-7} (R/r)^6$ dyne cm$^{-2}$ is small compared to
both the thermal pressures that we have computed at the planet surface
and the hydrodynamic pressures at the sonic point. For planetary
magnetic fields to confine the wind, they are required to have surface
strengths in excess of $\sim 30$ Gauss.

\subsection{\revised{Occurrence Rates of Close-In Progenitors}}
\label{sec_occurrence}
 
\revised{According to our analysis in \S\ref{sec_apply}, currently KIC
  1255b has a mass of at most $0.02$--$0.07 \, M\sub{\earth}$, and is
  in a final, short-lived, catastrophically evaporating phase possibly
  lasting another $t\sub{evap} \sim 40$--400 Myr during which dust is
  ejected at a large enough rate to produce eclipse depths of order
  $1\%$. But for most of its presumably Gyrs-long life, KIC 1255b was
  a more quiescent planet --- larger but still less than roughly
  Mercury in size (\S\ref{sec_apply}), with a stronger gravity and
  emitting a much more tenuous wind. We therefore expect that for every
  KIC 1255b-like object discovered, there are many more progenitors in
  the quiescent phase --- i.e., planets having sizes up to about that
  of Mercury orbiting main sequence K-type stars at $\sim$0.01 AU.}

\revised{
Out of the $\sim$45,000 main sequence K-type stars in the
Kepler Input Catalogue \citep{Batalha:2010p16185}, there is apparently
only 1 object like KIC 1255b. 
Then 
\begin{equation}
\label{eq_f}
	f\sub{observed}\sim f\sub{progenitor} f\sub{transit} f\sub{evap} \sim \frac{1}{45000},
\end{equation}
where $f\sub{progenitor}$ is the intrinsic occurrence rate of
progenitors around K stars with orbital periods shorter than a day,
$f\sub{transit}\sim R_{*}/a \sim 1/4$ is the geometric probability of
transit, and $f\sub{evap} = t\sub{evap}/t\sub{life} \sim \{40,400\} \,
\rm{Myr}\, /\, 5\, \rm{Gyr} \sim \{0.8,8\}\%$ is the fraction of the planet's
lifetime spent in the catastrophic mass-loss stage.
Inverting equation (\ref{eq_f}), we estimate that $f\sub{progenitor}
\sim \{1,0.1\}\%$ of K stars harbor a close-in planet having less than the mass of Mercury.}

\revised{For each KIC 1255b-like object there should be
  $f\sub{evap}^{-1} = t\sub{life}/t\sub{evap} \sim \{130,13\}$ planets
  with radii no larger than about Mercury's ($\lesssim 0.4
  \,R\sub{\earth}$), transiting K-type stars with sub-day
  periods. These planets have transit depths of $(R/R_{*})^2 \lesssim
  10^{-5}$ --- small but possibly detectable by \textit{Kepler} if
  light curves are folded over enough periods (S.~Rappaport 2012,
  personal communication) and if $\dot{M} \sim \max \dot{M}$ so that
  the progenitor masses attain their maximum, Mercury-like values. If $\dot{M}$
  were actually $\ll \max \dot{M}$, the progenitor sizes would be less
  than that of Mercury. The smaller sizes would decrease their lifetimes $t_{\rm
    evap}$ and thus increase their expected number $f\sub{evap}^{-1}$,
  but would also render them undetectable even with
  \textit{Kepler}.}

\subsection{Iron Planets}
\label{sec_iron}

\begin{figure}%
\includegraphics[width=\linewidth]{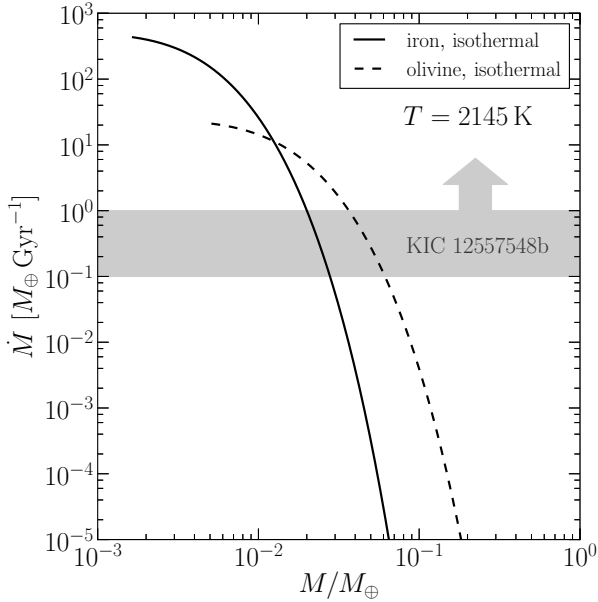}
\caption{%
  Mass loss rates $\dot{M}$, computed using the isothermal model at $T
  = 2145$ K, for iron planets and olivine planets of varying $M$. In
  the low-$M$, free-streaming limit, the higher vapor pressure of iron
  allows for a higher $\dot{M}$ than for olivine. At larger $M$, mass
  loss rates are lower for iron planets than for olivine planets
  because of the higher molecular weight of iron. Our estimate for the
  \revised{maximum} present-day mass of KIC 1255b varies by a factor of two between the
  iron and olivine scenarios.}
\label{fig_H}
\end{figure}

In our model, we found the \revised{(maximum)} present-day mass of KIC
1255b to be about 1/3 of its \revised{(maximum)} mass at formation. If KIC
1255b began its life similar in composition to the rocky planets in
the Solar System, evaporation may have stripped the planet 
of its silicate mantle, so that only its iron core remains: KIC 1255b
could be an evaporating iron planet today.


We estimate mass loss rates of a pure iron planet using our isothermal model, including tidal gravity. We set the mean molecular weight of the gas to $\mu\sub{Fe}\approx 56 m\sub{H}$ and use a bulk density for the planet of 8.0 g cm$^{-3}$, appropriate for a pure iron planet of mass $0.01 M\sub{\earth}$ \citep{Fortney:2007p16206}. We fit laboratory measurements of the iron vapor pressure\footnote{We have found measurements of the iron vapor pressure at $T\sim 2000$ K to differ by a factor of two in the literature (see, e.g., \citealt{Nuth:2003p16522}).} at $T = 2200$ K \citep{Desai:1986p16560} to equation (\ref{eq_pvap}), obtaining $e^{b}=7.8\e{11}$ dyne cm$^{-2}$ for a latent heat of sublimation $L\sub{sub}=6.3\e{10}$ erg g$^{-1}$ \citep{Desai:1986p16560} and $m=\mu\sub{Fe}$.

\begin{figure}%
\centering
\includegraphics[width=0.7\linewidth]{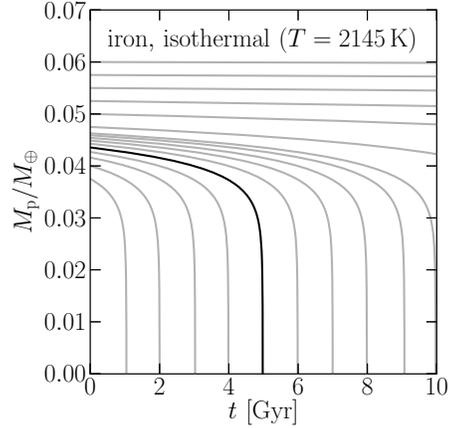}
\caption{
Mass-loss histories $M(t)$ for an iron planet, \revised{obtained by time-integrating $f\sub{duty} \cdot \max \dot{M}$ with $f\sub{duty} = 0.5$} for an isothermal wind. An iron planet with a 5 Gyr lifetime will have an initial mass of $\sim$0.044 $M\sub{\earth}$. As was the case for olivine planets (see Figure \ref{fig_G}), the catastrophic evaporation stage lasts only for $\sim$100 Myr. Iron planets (or planetary iron cores) with masses greater than $0.05 \,M\sub{\earth}$ survive for over 10 Gyr.}
\label{fig_I}
\end{figure}

In Figure \ref{fig_H} we compare mass loss rates derived for the
isothermal model with $T=2145$ K for both an iron planet and a pure
olivine planet.  On the one hand, at this temperature, the vapor
pressure for iron $P\sub{vapor,\,Fe}=1.8\e{3}$ dyne cm$^{-2}$ is about
50 times higher than for olivine, which raises the mass loss rate by a
similar factor in the free-streaming limit appropriate for low masses
(see equation \ref{eq_free}). On the other hand, at higher masses,
$\dot{M}$ for iron drops below that for olivine because the iron
atmosphere, with its higher molecular weight, is harder to blow off.
These two effects counteract each other so that an iron planet and a
silicate planet both reach $\dot{M}\sub{1255b} \sim 1 \,
M\sub{\earth}$ / Gyr at a similar $M$. Thus at least within the context of isothermal winds, our estimate for the present-day mass of KIC 1255b 
 is insensitive to whether the evaporating surface of
the planet is composed of iron or silicates.  Figure \ref{fig_I}
displays mass loss histories for an iron planet using our isothermal
model at $T=2145$ K. Initial planet masses are up to a factor of two lower
than those of our full model using silicates.

According to Figure \ref{fig_I}, iron planets with $M \gtrsim 0.05
M\sub{\earth}$ survive for tens of Gyrs. Compare this result to its
counterpart in Figure \ref{fig_G} (left panel), which shows that
olivine planets with $M\lesssim 0.1 M\sub{\earth}$ evaporate within
$\sim$10 Gyr. This comparison suggests that for planets with the right
proportion of silicates in the mantle to iron in the core, mantles may
be completely vaporized, leaving behind essentially non-evaporating
iron cores. Such massive, quiescent iron cores might be detectable by
\textit{Kepler} via direct transits (see \S\ref{sec_occurrence}).

\subsection{Dust Condensation}
\label{sec_condensation}

Recall that we have not modelled the microphysics of dust formation,
but treated the dust-to-gas ratio as a free function. Are the
conditions of our flow actually favorable for the condensation of
dust? \rev{Any gas whose partial pressure is greater than its vapor
  pressure (at that $T$) might condense and form droplets (modulo the
  many complications discussed below).} Condensation \rev{might}
proceed until all vapor in excess of saturation is in cloud
particles. If the condensates have sedimentation velocities larger
than updraft speeds they will rain out of the atmosphere; otherwise
they remain aloft (see, e.g., \citealt{Lewis:1969p16144,
  Marley:1999p16150, Ackerman:2001p16041, Helling:2001p17233}).

\begin{figure}%
\centering
\includegraphics[width=0.9\linewidth]{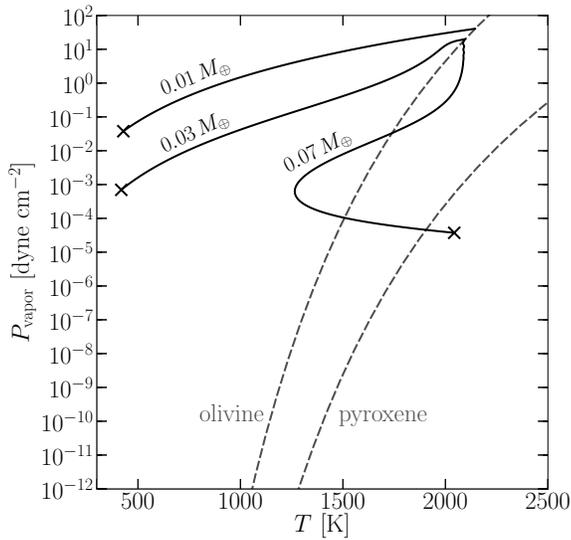}
\caption{%
Trajectories of the wind in pressure-temperature space, computed using the full model for $M=0.01$, $0.03$, and $0.07M\sub{\earth}$. Gas pressures generally remain above the equilibrium vapor pressures of pyroxene, permitting the condensation of pyroxene grains.}
\label{fig_K}
\end{figure}

In Figure \ref{fig_K}, we compare gas pressures of our solutions 
(for $\dot{M} = \max \dot{M}$) at
$M=0.01$, $0.03$, and $0.07\,M\sub{\earth}$ with the equilibrium vapor
pressures of olivine and pyroxene. At the base of the flow, the gas
pressure equals the saturation vapor pressure of olivine by
construction. As gas expands and cools, its pressure generally remains
above the saturation pressure of pyroxene, so conditions are favorable
for the condensation of pyroxene grains. Unfortunately we have an
embarrassment of riches: at $T \lesssim 1700$ K, our gas pressures are
many orders of magnitude higher than silicate vapor pressures, and the
concern is that gas condensation will lower pressures precipitously to
the point where winds shut down. This fate could be avoided if grains
take too long to condense out of the outflowing and rapidly rarefying
gas --- i.e., the finite timescale of grain condensation might permit
gas to remain supersaturated. \revised{Furthermore, even before bulk
  condensation can begin, seed particles must first nucleate from the
  gas phase. These issues call for a time-dependent analysis of grain formation, perhaps along
 the lines made for brown dwarf and exoplanet atmospheres (\citealt{Helling:2001p17233, Helling:2008p17236, Helling:2008p17230, Helling:2009p17239}).
 Accounting for the extra heating from small grains with sizes $\ll 1$ $\mu$m as they first
 condense out of the wind (\S\ref{sec_1fluid}) would keep the gas closer
 to isothermal and help to prevent catastrophic condensation.}

  Of course, by the wind leaves the sonic point and is diverted into
  the trailing comet-like tail, it must be full of grains to explain
  the observed occultations of KIC 1255b. In this paper, we have not
  modeled the dusty tail at all. Future work should address the
  dynamics of the tail and compute its extinction profile, with the
  aim of reproducing the observed transit lightcurves.

\section{CONCLUSIONS}
\label{sec_conclusions}

Our work supports the hypothesis that the observed occultations of the K-star KIC 12557548 originate from a dusty wind emitted by an evaporating planetary companion (``KIC 1255b''). We reach the following conclusions.

\begin{enumerate}

\item{\textit{Maximum present-day mass}}

  \revised{Our estimates for the maximum present-day mass of KIC 1255b
    range from $\max M \approx 0.02 M\sub{\earth}$ (roughly twice the
    mass of the Moon) to $\max M \approx 0.07 M\sub{\earth}$ (slightly
    larger than the mass of Mercury), depending on how strongly the
    wind is heated by dust and can remain isothermal (Figure
    \ref{fig_F}). For these maximum masses, calculated mass loss rates
    peak at $\max \dot{M} \sim 0.1$ $M\sub{\earth}$/Gyr, which is just
    large enough to produce the observed transit depths of
    order 1\%.  Smaller planets with weaker gravities yield larger
    mass loss rates and are also compatible with the observations.}

\item{\textit{Maximum formation mass}}

  \revised{For an assumed planet age of $\sim$5 Gyr, the maximum mass
    of KIC 1255b at formation ranges from $0.06 \,M\sub{\earth}$ to 0.1
    $M\sub{\earth}$, again depending on the energetics of the wind
    (Figure \ref{fig_G}).}

\item{\textit{Mass threshold for catastrophic evaporation}}

A pure rock planet of mass $\gtrsim 0.1 M\sub{\earth}$ and surface temperature $\lesssim 2200$ K will survive with negligible mass loss for tens of Gyrs.  

\item{\textit{Time variability}}

  The observed occultations of KIC 12557548 vary by up to an order of
  magnitude in depth without any apparent correlation between
  orbits. The implied order-of-magnitude variations in $\dot{M}$ can
  be explained in principle by the exponential sensitivity of
  $\dot{M}$ to conditions at the planet surface. The dynamical
  ``refresh'' timescale of the wind $t\sub{dyn} \sim 14$ h is similar
  to the orbital period $P\sub{orb}=15.7 $ h. This supports our dusty wind model because were $t\sub{dyn} \gg
  P\sub{orb}$ or $t\sub{dyn} \ll P\sub{orb}$, then eclipse depths
  would correlate from orbit to orbit, in violation of the
  observations.

\item{\textit{Progenitors and occurrence rates}}

  \revised{KIC 1255b's current catastrophic mass loss phase may represent
    only the final few percent of the planet's life. As such, for
    every KIC 1255b-like object, there could be anywhere from 10 to
    100 larger planets in earlier stages of mass loss. These close-in,
    relatively quiescent progenitors may be detectable by
    \textit{Kepler} through conventional ``hard-sphere'' transits if
    they are as large as Mercury. We cannot, however, rule out the
    possibility that the progenitors are lunar-sized or even
    smaller. If KIC 1255b remains the only catastrophically
    evaporating planet in the \textit{Kepler} database, then the
    occurrence rate of close-in progenitors orbiting
    K-stars with sub-day periods is $> 0.1\%$, with larger occurrence rates
    for progenitors increasingly smaller than Mercury.}

\item{\textit{Iron planet}}

  KIC 1255b may have lost $\sim$70\% of its formation mass to its thermal
  wind. It seems possible that today only the iron
  core of KIC 1255b remains and is evaporating. Transmission spectra
  of the occulting dust cloud might reveal whether the planet's
  surface is composed primarily of iron or silicates.

\item{\textit{Future modelling and observations}}

  Keeping the wind hot as it lifts off the planet surface
  significantly enhances mass loss rates. In our model we have
  included heating of the gas by micron-sized grains embedded in the
  flow. But these grains can also shadow the surface from starlight
  and reduce $\dot{M}$ if they are too abundant. Additional heating
  may be provided by super-blackbody grains with sizes $\ll 1$ $\mu$m
  which do not significantly attenuate starlight. Future models should
  incorporate such tiny condensates --- and treat the dusty comet-like
  tail that our paper has completely ignored. More information about
  grain size distributions and compositions might be revealed by
  \textit{Hubble} observations of KIC 12557548 scheduled for early
  2013.

\end{enumerate}

\section{ACKNOWLEDGMENTS}

We thank Saul Rappaport for alerting us to the feasibility of
detecting Mercury-sized progenitors of KIC 1255b using
\textit{Kepler}, and Andrew Howard and Geoff Marcy for sharing their
Keck observations of KIC 12557548.  We are also grateful to Andrew
Ackerman, Tom Barclay, Bill Borucki, Raymond Jeanloz, Hiroshi Kimura,
Edwin Kite, Michael Manga, Mark Marley, and Subu Mohanty for useful
and encouraging conversations; \revised{to Paul Kalas for pointing us
  to work on beta Pictoris that presaged the story of KIC 1255b; and
  to an anonymous referee for a helpful report that motivated us to
  review the various modeling uncertainties and introduced us to the
  cloud formation literature.}  D.P.-B.~acknowledges the support of a
UC MEXUS-CONACyT Fellowship.  E.C.~acknowledges \textit{Hubble Space
  Telescope} grant HST-AR-12823.01-A.

\bibliographystyle{aa}
\bibliography{adsdpb.bib}

\end{document}